\documentclass[twocolumn]{article}
\def\thebibliography#1{\list{~\arabic{enumi}.}
  {\settowidth\labelwidth{#1.}\leftmargin=2.47em
   \labelsep\leftmargin \advance\labelsep-\labelwidth
   \usecounter{enumi}}\def\makelabel##1{\rlap{##1}\hss}%
   \def\newblock{\hskip 0.11em plus 0.33em minus -0.07em}
   \sloppy \clubpenalty=4000 \widowpenalty=4000 \sfcode`\.=1000\relax}

\newcounter{tempref}
\begin{document} 
\title{Resource Letter: Bio-molecular Nano-machines:\\ where Physics, Chemistry, Biology and Technology meet}
\author{Debashish Chowdhury{\footnote{E-mail: debch@iitk.ac.in}},\\
Department of Physics,\\
Indian Institute of Technology,\\
        Kanpur 208016, India}
\maketitle
\begin{abstract}
Cell is the structural and functional unit of life.
This Resource Letter serves as a guide to the literature on nano-machines 
which drive not only intracellular movements, but also motility of the 
cell. These machines are usually proteins or macromolecular assemblies 
which require appropriate fuel for their operations. Although, traditionally, 
these machines were subjects of investigation in biology and biochemistry, 
increasing use of the concepts and techniques of physics in recent years 
have contributed to the quantitative understanding of the fundamental 
principles underlying their operational mechanisms. The possibility of 
exploiting these principles for the design and control of artificial 
nano-machines has opened up a new frontier in the bottom-up approach 
to nano-technology.
\end{abstract}


{\it {Some are to be read, some to be studied, and some may be neglected 
entirely, not only without detriment, but with advantage. } - Anonymous}\\


\section{Introduction}

Motility is the hallmark of life. From the sprinting leopard to flying 
birds and swimming fish, movement is one of life's central attributes. 
The mechanisms of motility at the level of macroscopically large 
organisms are the main topics of investigation in biomechanics and 
insights gained from these investigations find applications, for 
example, in robotics. Not only animals, but even plants also move in 
response to external stimuli. Results of pioneering systematic study of 
this phenomenon were reported already in the nineteenth century by 
Charles Darwin in a classic book, titled {\it The power of movement in 
plants}, which was co-authored by his son. 

In living systems, movements take place at all levels of biological 
organization- from molecular movements at the subcellular levels and 
cellular movements to movements of organs and organ systems. However, 
in this article, we focus exclusively on the molecular mechanisms of 
motility at the level of single cells (both unicellular organisms and 
individual cells of multicellular organisms) and those at the 
subcellular level.



\subsection{Cell movements: molecular mechanisms of motility}

Antonie van Leeuwenhoek made the first systematic study of the motility 
of unicellular microorganisms using his primitive microscope. Since 
then, over the last three centuries, swimming, crawling, gliding and 
twitching of single cells have fascinated generations of biologists. 
However, investigation of the molecular mechanisms of cellular motility 
began only a few decades ago. The motility of a cell is the outcome of  
the coordination of many intracellular dynamical processes. Interestingly, 
intracellular movements also drive motility and division of the cell 
itself. We'll present a systematic list of these developments from 
the perspective of physicists.



\subsection{Intracellular movements: machines and mechanisms}

\noindent {\it ``Nature, in order to carry out the marvelous operations
in animals and plants, has been pleased to construct their organized
bodies with a very large number of machines, which are of necessity
made up of extremely minute parts so shaped and situated, such as to
form a marvelous organ, the composition of which are usually invisible
to the naked eye, without the aid of the microscope''}. \\
Marcelo Malpighi, 17th century (as quoted by Marco Piccolino,  
Nat. Rev. Mol. Cell Biol. 1, 149-153 (2000)). 

Imagine an under water ``metro city'' which is, however, only about 
$10 \mu$m long in each direction! In this city, there are ``highways'' 
and ``railroad'' tracks on which motorized ``vehicles'' transport 
cargo to various destinations. It has an elaborate mechanism of 
preserving the integrity of the chemically encoded blueprint of the 
construction and maintenance of the city. The ``factories'' not only 
supply their products for the construction and repair works, but 
also manufacture the components of the machines. This eco-friendly 
city re-charges spent ``chemical fuel'' in uniquely designed ``power 
plants''. This city also uses a few ``alternative energy'' sources 
in some operations. Finally, it has special ``waste-disposal plants'' 
which degrade waste into products that are recycled as raw materials 
for fresh synthesis. This is not the plot of a science fiction, but 
a dramatized picture of the dynamic interior of a cell.

In an influential paper, published in 1998, Bruce Alberts  
emphasized that ``the entire cell can be viewed as a factory that 
contains an elaborate network of interlocking assembly lines, each 
of which is composed of a set of large protein machines''.  Just like 
their macroscopic counterparts, molecular machines have an ``engine'', 
an input and an output. Some of these machines are analogous to motors 
whereas some others are like pumps; both linear and rotary motors 
have been identified. Some motors move on protein filaments whereas 
others move on nucleic acid strands (i.e., DNA or RNA). \\



\noindent {\it ``Where bacillus lives, gravitation is forgotten, and
the viscosity of the liquid, the resistance defined by Stokes' law,
the molecular shocks of the Brownian movement, doubtless also the
electric charges of the ionized medium, make up the physical environment
and have their potent and immediate influence on the organism. The
predominant factors are no longer those of our scale; we have come to
the edge of a world of which we have no experience, and where all our
preconceptions must be recast''}. \\
- D'Arcy Thompson, in {\it On Growth and Form}, vol.I reprinted
2nd edition (Cambridge University Press, 1963). \\

In spite of the striking similarities, it is the differences between 
molecular machines and their macroscopic counterparts that makes the 
studies of these systems so interesting from the perspective of 
physicists. Biomolecular machines are usually protein or macromolecular 
complex. These operate in a domain far from thermodynamic equilibrium 
where the appropriate units of length, time, force and energy are, 
{\it nano-meter}, {\it milli-second}, {\it pico-Newton} and $k_BT$, 
respectively ($k_B$ being the Boltzmann constant and $T$ is the 
absolute temperature). The viscous forces and random thermal forces on 
a nano-machine dominate over the inertial forces. These are made of 
soft matter and are driven by ``isothermal'' engines. Molecular motors 
can convert chemical energy directly into mechanical energy.



\subsection{Outline of organization}

We divide the intracellular molecular cargoes into three different types: 
(i) membrane-bound cargoes, e.g., vesicles and organelles; 
(ii) macromolecules, e.g., DNA, RNA and proteins; 
(iii) medium-size organic molecules and small inorganic ions. 
In part I we study motor proteins which transport the membrane-bound 
cargoes.  In part II we consider all those machines which are involved 
in the synthesis, export/import, packaging, other kinds of manipulations 
and degradation of the macromolecules. In part III we focus on machines 
which transport medium-size organic molecules and small inorganic ions 
across plasma membrane or internal membranes of eukaryotic cells; 
transporters of ions are usually referred to as pumps because ions are 
transported against their natural electro-chemical gradients. Finally, 
in part IV we present machines and mechanisms which drive cell motility 
and cell division.

Based on the nature of input and output energies, machines can be 
classified. For example, the motor of hair dryer is an 
electro-mechanical machine. But, in this article we'll not consider 
purely chemo-chemical machines although some of these perform 
important biological functions. 



\subsection{Criteria for selection}

We have used the following guidelines for selection of papers for this 
resource letter: \\

(i) To our knowledge, at present, there is no single book where a reader 
can find a comprehensive coverage of all the molecular machines. 
Therefore, in this resource letter, we list monographs and edited 
collections of reviews on specific machines and mechanisms. \\
(ii) Review articles usually provide a critical overview of progress in 
an area of research and, normally, remain useful to both beginners as 
well as experts for a relatively longer period of time as compared to 
original papers. Therefore, in this resource letter, review articles 
guide the reader through the enormous literature on {\it experimental} 
works on molecular machines. Occasionally, we also list original 
experimental papers; most of these are either classic or too recent to 
be discussed in any review article, or introduce new models.\\ 
(iii) Since the emphasis of this resource letter is on {\it quantitative} 
models of mechanisms of molecular machines, many original papers on 
{\it theoretical} works have been listed together with the review articles. 
(iv) Results of fundamental research on the structure and function of 
molecular machines not only have important biomedical implications but 
may also find practical applications in bottom-up approach to designing 
and manufacturing artificial nano-machines. Therefore, papers on 
bio-nanotechnology which satisfy the criteria (ii) or (iii) above have 
also been listed. 
(v) Unpublished manuscripts (including those posted in public domain 
archives) have not been listed. But, the final version of some Ph.D. 
theses have been included because these provide technical details 
which are not available in the papers published elsewhere by the author.

\subsection{List of review series and journals}

In this multidisciplinary area of research, articles appear in journals 
that cover physics, chemistry, biology and (nano-)technology. We list 
here only a few major sources for review articles as well as original 
papers. But, this list is neither exhautive nor in the order of any 
ranking.\\

(1) ``Annual Review'' Series (e.g. Annual Reviews of Biophysics and 
Biomolecular Structure). \\
(2) ``Trends'' series (e.g., Trends in Cell Biology). \\
(3) ``Current Opinion'' series (e.g., Current Opinion in Structural Biology).\\
(4) Bioessays,\\
(5) ``Nature Reviews'' series (e.g., Nature Reviews in Microbiology).\\ 
(6) Nature, \\
(7) Nature Cell Biology,\\
(8) Nature Structural and Molecular Biology,\\
(9) Science, \\
(10) Proceedings of the National Academy of Sciences, USA, (PNAS),\\
(11) Cell,\\
(12) Molecular Cell,\\
(13) Current Biology.\\
(14) Journal of Molecular Biology,\\
(15) Journl of Cell Biology, \\
(16) Journal of Biological Chemistry,\\
(17) Biophysical Journal,\\
(18) Physical Review Letters,\\
(19) Physical Review E,\\
(20) Physical Biology,\\
(21) Europhysics Letters,\\
(22) European Physical Journal E.\\
(23) EMBO Reports,\\
(24) EMBO Journal,\\
(25) European Biophysical Journal.\\

\section{Experimental techniques for studying operational mechanisms of molecular machines}

{\it ``The most profound scientific revolutions are those that provide
an entirely new way of viewing and studying a field. These are the ones
that provoke new questions and question old answers and in the end give
us a new understanding of what we thought we understood. Often they are
occasioned by the invention of novel instruments of techniques; the
telescope, the microscope, and X-ray diffraction come to mind. It may
be that such a revolution is occurring in biochemistry today through
the development of methods that allow us to investigate the dynamics
of single macromolecules in real time.''}- K.E. van Holde, J. Biol. Chem.
{\bf 274}, 14515 (1999).

Seeing is believing. Telescopes opened up the celestial world in front
of our eyes. The invention of the optical microscopes in the seventeenth
century made it possible to have a glimpse of the world of micro-organisms
(bacteria, etc.). But these microbes are typically micron-size objects;
it would be ideal if we could have ``nanoscopes'' for seeing nano-machines.
In addition to the requirement of high spatial resolution, such nanoscopes
should also have sufficiently high temporal speed so that the dynamics of
the nano-machines can be monitored under the nanoscope.

But, it is impossible to see a molecule directly under an optical microscope
of conventional design because nature has imposed a limit on the resolution
that can be acheived with these optical instruments. This fundamental limit
on the resolution is a consequence of the wave nature of light and it
depends on the wavelength of the radiation used for observation.

But, optical microscopes merely enhance the power of our visionary
perception. Therefore, in principle, it should be possible to achieve
higher resolution if X-rays or $\gamma$-rays are used for imaging
although we can no lnger use our eyes as dectector for these probes.
Moreover, vision is only one of the several sensory perceptions humans
possess. A blind person can construct a mental image of an object by
running his fingers along the contours of the object. Furthermore, in
principle, it is possible to reconstruct the shape of an object, without
seeing or touching it, by throwing balls at it from all sides and, then,
analyzing the way the balls are scattered by the object.

\subsection{Ensemble-averaged techniques}

$\bullet${\bf X-ray crystallography and electron microscopy}

The basic principle of X-ray scattering for the determination of the
strcture of macromolecules is as follows:
an atomic constituent of the macromolecule absorbs some energy of the
X-ray incident on it and then re-radiates the same in all directions.
A protein crystal has a periodic array of identical atoms. The X-rays
re-radiated by these atoms interfere constructively in some directions
whereas they interfere destructively in all the other directions.
Therefore, the detectors record a ``pattern'' in the intensity of X-ray
scattered by the protein crystal sample. But, such a  ``diffraction
pattern'' provides an indirect, and static, image of a molecular machine.
However, microscopes (optical as well as electron) have some advantages
over the X-ray scattering technique:
microscopes produce the images directly in real space whereas X-ray
diffraction requires Fourier transform from momentum space to real space.

The deBroglie wavelength associated with a material particle is given by
$\lambda = h/p$ where $p$ is the momentum of the particle. A sufficiently
high resolution microscope can be constructed if a charged particle is
selected and it is accelerated to the required momentum by applying an
external electric field. Electrons are most convenient for this purpose;
an electron beam can be easily bent and focussed using a suitable magnetic
field configuration. Electron microscopy is one of the most powerful
experimental techniques for determination of the structures of molecular
machines. 
In spite of all the technological advances in electron microscopy, it
is still lot more cumbersome to use than an optical microscope. In an
optical microscope, the sample does not require as elaborate preparation
as in an electron microscope. Besides, the intense beam of electrons
often damage or destroy the sample itself. Moreover, image obtained from
an electron microscope requires special expertise to interpret.
Furthermore, the generation and control of the electro-magnetic fields
makes the electrom microscope costly as well as much less user friendly
than optical microscopes.



\subsection{Single-molecule techniques}

We got our first glimpse of the macromolecules via X-ray diffraction
and, then, electron microscopy. But, what one got from those probes
were static pictures. Moreover, most of the traditional old experimental
techniques of biophysics relied on collection of data for a large
collection of molecules and thereby getting their ensemble-averaged
properties. The amplification of the signals caused by the presence of
large number of such molecules makes it easier to detect and collect
the data. The average value of a variable is valuable information.
There are practical limitations of the bulk measurements in the
specific context of understanding the operational mechanisms of
cyclic molecular machines because it is practically impossible to
synchronize their cycles. The recently developed single-molecule 
techniques can be broadly
classified into two groups: (i) methods of imaging, and (ii) methods
of manipulation.



$\bullet$ {\bf Techniques of single-molecule imaging}

For visualization of
the conformational changes or movements of the molecule under
investigation in a single molecule experiment, a prior attachment of
a label to the molecule is essential. Based on these labels, the 
single molecule imaging of molecular motors can be divided into two
groups: (i) techniques where the label is a relatively large 
light-scattering object (for example, a dielectric bead of $1$ micron 
diameter); and (ii) techniques where the label itself emits light 
(e.g., a fluorophore). Fluorescence microscopy provided a glimpse 
(howsoever hazy) of single molecules. Imaging a fluorescently labelled 
molecular motor in real time enables us to study its dynamics just 
as ecologists use ``radio collars'' to track individual animals.



$\bullet$ {\bf Techniques of single-molecule manipulation: tweezers and AFM} 

Let us also not forget that experiments are also carried out with
molecular machines to undestand their mechanisms; such experients should
answer questions like ``what if...''. Such controlled experiments would
need some means of manipulating the nanomachines of life.
The mechanical transducers like, for example, cantilevers of scanning
force microscopes (SFM) and microneedles, require physical contact with
the biomolecule. In contrast, manipulators that utilize electromagnetic
fields do not require any contact forces.

The operation of optical tweezers is based on a very simple physical
principle. Photons are massless (more precisely, rest mass is zero),
but have momentum. Photons are capable of exerting very weak force
called radiation pressure in the terminology of classical physics.
Utilizing this property of photons (or, equivalently, electromagnetic
radiation) in a laser beam with high, but inhomogeneous, intensity,
it has been posible to trap dielectric particles (e.g., a latex bead)
at the focal point of the beam. When a motor attached to such a 
bead walks on its filamentary track, the laser trap pulls it back
thereby applying a load force on the motor.

In magnetic tweezers, the macromolecule is attached between a
surface and a superparamagnetic bead. Stretching force can be applied on
the macromolecule by controlled alterations of the external magnetic field.
A major advantage of the magnetic tweezer is that the same set up can be
used also to apply torque on the molecule by merely rotating the magnetic
field.



\subsection{Experimental model systems}

From the evolutionary point of view, cells can be broadly divided into
two categories, viz., {\it prokaryotes} and {\it eukaryotes}. Most of
the common bacteria (like, for example, Escherichia Coli and Salmonella)
are prokaryotes. Animals, plants and fungi are collectivelly called
eukaryotes. The main difference between prokaryotic and eukaryotic 
cells lies in their internal architectures; the main distinct feature 
of eukaryotic cells is the cell nucleus where the genetic materials are 
stored. The prokaryotes are mainly uni-cellular organisms. The eukaryotes 
which emerged first through Darwinian evolution of prokaryotes were 
also uni-cellular; multi-cellular eukaryotes appeared much later. 

In biology, often the simplest among a family of objects is called a 
model system for the purpose of experimental investigations. \\

$\bullet$ {\bf Model eukaryotes}:

The most popular model {\it animals} for biological studies are as 
follows: 
(i) the fruit fly {\it Drosophila melanogaster}, a model insect,
(ii) {\it Caenorhabditis elegans} (C-elegans), a transparent worm, 
(iii) the zebra fish {\it danio rerio}, a model vertebrate;  
(iv) the mouse, however, is more important for practical use of cell 
biology in medical sciences.
{\it Arabidopsis thaliana} is the most popular model {\it plant} while 
{\it Chlamydomonas reinhardtii} is a model of green algae. {\it 
Saccharomyces cerevisiae} (Baker's yeast) and {\it Schizosaccharomyces 
pombe} (Fission yeast) are most widely used models for fungi. However, 
for studying filamentous fungi, {\it Neospora crassa} is used most 
often as a model system.\\

$\bullet$ {\bf Model prokaryotes}:

Bacteria are divided into two separate groups on the basis of their 
response to a staining test invented by Hans Christian Gram. Those 
which respond positively are called Gram-positive bacteria whereas 
those whose response is negative are called Gram-negative. One of 
the main differences between these two groups of bacteria is the 
nature of the cell wall.

The commonly used models for Gram-positive bacteria are {\it Bacillus 
subtilis}, {\it Listeria monocytogenes}, etc. The bacterium {\it 
Escherichia coli} (E-coli), which is normally found in the colon of 
humans and other mammals, and the bacterium {\it Salmonella} are the 
most extensively used model for Gram-negative bacteria. Another prominent 
member of the group of Gram-negative bacteria is {\it Proteus mirabilis}.\\ 

$\bullet$ {\bf Model viruses and bacteriophages}:

Human immunodeficiency virus (HIV) is the most dreaded among the viruses
that can infect {\it homo sapiens} (humans). Among the other viruses 
which can infect eukaryotes are {\it Tobacco mosaic virus }, etc. 
Bacteriophages are also viruses, but these infect prokaryotes.
T-odd (e.g., T7) and T-even (e.g., T4) bacteriophages, phage $\lambda$, 
$\phi29$, etc. are some of the extensively used model bacteriophages.



\section{Techniques for theoretical modeling of molecular machines}

Theory provides {\it understanding} and {\it insight}. These allow
us not only to interpret the empirical observations and recognize
the importance of the various ingredients but also to generalize,
to create a framework for addressing the next level of question and
to make predictions which can be tested in in-vivo/ in-vitro or
in-silico experiments.

Theorization requires a model of the system. A theoretical model is 
an abstract representation of the real system which helps in 
understanding the real system. This representation can be pictorial 
(for example, in terms of cartoons or graphs) or symbolical (e.g., 
a mathematical model). Qualitative predictions may be adequate for 
understanding some complex phenomena or for ruling out some plausible 
scenarios. But, a desirable feature of any theoretical model is that 
it should make quantitative predictions. The predictions of a theory, 
at least in principle, can be tested by in-vitro and/or in-vivo 
experiments in the laboratory.

The predictions of a mathematical model can be derived {\it analytically} 
in terms of abstract symbols; for specific sets of values of the model 
parameters, the predictions can be shown numerically or graphically. 
The predictions of a theoretical model can be obtained {\it numerically} 
by carrying out computer simulations (i.e., {\it in-silico} experiments)   
of the model. Thus, simulation is not synonymous with modeling. When a 
model is too complicated to be formulated in abstract notations and to 
be treated analytically, it is called a computer model of the system. 
Since fully analytical treatment of a model can be accomplished
exactly only in rare cases, one has to make sensible approximations
so as to get results as accurate as possible. Simulation of a model 
also tests the validity of the approximations made in the analytical 
treatments of the model. We should also make a distinction between 
the two different ``computational methods'', namely, (i) computer 
simulations which, as we have mentioned above, test hypotheses; and 
(ii) {\it Knowledge discovery} (or, {\it data mining}) which extracts
hidden patterns or laws from huge quantities of experimental data,
forming hypotheses.

A model can be formulated at different physical or logical levels of 
resolution. The physical resolution can be spatial resolution or 
temporal resolution. Every theoretical model is intended to address 
a set of questions. The modeler must choose a {\it level of description} 
appropriate for this purpose keeping in mind the phenomena that are 
subject of the investigation. Otherwise, the model may have either 
too much redundant details or it may be too coarse to provide any useful 
insight. Since physicists most often focus only on generic features of the
various classes of machines, rather than specific features of individual
members of these classes, they normally develop minimal models which may
be regarded as {\it mesoscopic}, rather than molecular, i.e., their status
in somewhere in between those of the macroscopic and molecular models.



\subsection{Mechanics of molecular machines: noisy power stroke versus Brownian ratchet}

If the input energy directly causes a conformational change of the  
protein machiney which manifests itself a mechanical stroke of the 
machine, the operation of the machine is said to be driven by a 
``power stroke'' mechanism. This is also the mechanism used by all 
man made macroscopic machines. Let us contrast this with the following 
alternative scenario: suppose, the machine exihibits ``forward'' and 
``backward'' movements because of spontaneous thermal fluctuations. 
If now energy input is utilized to prevent ``backward'' movements, 
but allow the ``forward'' movements, the system will exhibit 
directed, albeit noisy, movement in the ``forward'' direction. 
Note that the forward movements in this case are caused directly by 
the spontaneous thermal fluctuations, the input energy rectifies the 
``backward'' movements. This alternative scenario is called the 
Brownian ratchet mechanism. 



\subsection{Chemical reactions relevant for molecular machines: ATP hydrolysis}

To understand molecular
machines, we also have to consider {\it chemical reactions}, which most
often supply the (free-) energy required to drive these machines. In other
words, in order to understand the mechanisms of biomolecular machines,
it is necessary to understand not only how these move in response to the
mechanical forces but also how these are affected by chemical reactions.
In fact, the machines are usually enzymes (i.e., catalysts). 

Adenosine triphosphate (ATP) contains three phosphate groups as compared 
to two phosphate groups in the adenosine diphosphate (ADP). 
Hydrolysis of ATP to ADP releases free energy and,
therefore, plays a crucial role in running a wide range of chemical
processes in a living organism. Therefore, ATP is sometimes also 
referred to as the ``energy currency'' of the cell.



\subsection{General mechano-chemistry of molecular machines} 

From the perspective of (bio-)physics, the mechanical force required for 
the directed movement of the motors are generated, most often, from the 
energy liberated in chemical reactions, e.g., in ATP hydrolysis. On the 
other hand, from the perspective of (bio-)chemistry, most of the machines 
are enzymes (i.e., proteins which act as catalysts for many chemical 
reactions); the rate of enzymatic reactions, including that of ATP 
hydrolysis, is strongly influenced by external forces.  Thus, the 
mechanisms of molecular machines are governed by a nontrivial combination 
of the principles of nano-mechanics and those of chemical reactions. 
Therefore, quantitative modeling of molecular machines require theoretical 
formalisms of {\it mechano-chemistry} or {\it chemo-mechanics}. 



$\bullet${\bf Effect of force on chemical reactions: force-dependence of ATP hydrolysis}



$\bullet$ {\bf Hydrolysis and phosphorylation}



$\bullet$ {\bf Efficiency of molecular machines: general discussions}

The efficiency of molecular motors can be defined in several different  
ways: while one of the definitions is very similar to that of its 
macroscopic counterpart, the other definitions are unique to motors 
operating under different conditions and characterize different aspects 
of its movement.

Not all molecular motors are designed to pull loads. Moreover, in
contrast to the macroscopic motors, viscous drag forces strongly 
influence the function of molecular motors. Therefore, there is a 
need for a generalized definition of efficiency that does not 
necessarily require the application of any external load force.
Such a measure of efficiency, which is different from the
thermodynamic efficiency defined above, has also been suggested;
it is called ``Stokes efficiency'' because the viscous drag is 
calculated from Stokes law.



$\bullet$ {\bf Allosterism and molecular motors}

Allosterism usually refers to the change of conformation around one 
location of a protein in response to binding of a ligand to another 
location of the same protein. A motor protein has separate sites for 
binding ATP and the track. Therefore, the mechanochemical cycle of a 
motor can be analyzed from the perspective of allosterism. 



\vspace{1cm} 
\noindent {\bf Part I: Cytoskeletal motors: porters and rowers, shuttles and muscles}\\

The cytoskeleton of an eukaryotic cell maintains its architecture. 
Counterparts of some molecular components of the eukaryotic 
cytoskeleton have been discovered recently also in prokaryotic cells. 
It is a complex dynamic network that can change in response to
external or internal signals. The cytoskeleton is also responsible 
for intra-cellular transport of packaged molecular cargoes as well 
as for the motility of the cell as a whole. The cytoskeleton plays 
crucially important role also in cell division and development of 
organisms. In this part we focus almost exclusively on the motility 
and contractility driven by molecular motors at the sub-cellular 
level; motor-driven motility and contractility of the cell as a 
whole will be taken up in the last part of this article.

\section{Eukaryotic cytoskeleton: structure and dynamics}

\subsection{Protein constituents of eukaryotic cytoskeleton}

The protein constituents of the cytoskeleton of eukaryotic cells
can be broadly divided into the following three categories: 
(i) {\it Filamentous} proteins, (ii) {\it accessory} proteins, and
(iii) {\it motor} proteins. 
The three classes of filamentous proteins, which form the main 
scaffolding of the cytoskeleton, are: 
(a) {\it actin}, (b) {\it microtubule}, and (c) {\it intermediate
filaments}.  
On the basis of functions, accessory proteins can be categorized 
as follows: 
(i) regulators of filament polymerization,
(ii) filament-filament linkers,
(iii) filament-plasma membrane linkers.
Since accessory proteins do not play any crucially important role in 
the operation of the cytoskeleton-based molecular machines, we shall 
not consider accessory proteins in this article.

The three superfamilies of motor proteins are: 
(i) {\it myosin} superfamily,
(ii) {\it kinesin} superfamily, and
(iii) {\it dynein} superfamily.
Both kinesins and dyneins move on microtubules; in contrast, myosins
either move on actin tracks or pull the actin filaments.\\

$\bullet$ {\bf Structures of microtubules and actin filaments}

Microtubules are cylindrical hollow tubes whose diameter is approximately
20 nm. The basic constituent of microtubules are globular proteins called 
tubulin. Hetero-dimers, formed by $\alpha$ and $\beta$ tubulins, assemble 
sequentially to form a protofilament. 13 such protofilaments form a 
microtubule. The length of each $\alpha-\beta$ dimer is about 8 nm. 
Since there is only one binding site for a motor on each dimeric subunit 
of MT, the minimum step size for kinesins and dyneins is 8 nm.

Although the protofilaments are parallel to each other, there is
a small offset of about 0.92 nm between the dimers of the neighbouring
protofilaments. Thus, total offset accumulated over a single looping of
the 13 protofilaments is $13 \times 0.92 \simeq 12 nm$ which is equal
to the length of three $\alpha-\beta$ dimers joined sequentially.
Therefore, the cylindrical shell of a microtubule can be viewed as
{\it three} helices of monomers. Moreover,
the asymmetry of the hetero-dimeric building block and their parallel
head-to-tail organization in all the protofilaments gives rise to the
polar nature of the microtubules. The polarity of a microtubule is such
an $\alpha$ tubulin is located at its - end and a $\beta$ tubulin is
located at its + end. 

Filamentous actin are polymers of globular actin monomers. Each
actin filament can be viewed as a double-stranded, right handed
helix where each strand is a single protofilament consisting of
globular actin. The two constituent strands are half staggered
with respect to each other such that the repeat period is 72 nm.

The spatial organization and function of the cytoskeletons of plants 
and algae differ significantly from those of eukaryotic cells. 



$\bullet$ {\bf MAPs and ARPs}

Microtubule-associated proteins (MAPs) and Actin-related proteins 
(ARPs) play important roles in controlling the structure and dynamics 
of microtubules and filamentous actin, respectively. Microtubule 
plus-end tracking proteins (+TIPs) are special MAPs that accumulate 
at the plus end of microtubules; members of a few families of motor 
proteins are also +TIPs. Biological functions of some of these 
proteins will be considered later in this part of this resource 
letter.



\subsection{Prokaryotic cytoskeleton}

Unlike eukaryotes, bacteria do not have any obvious need for a cytoskeleton.
First, their cell walls are rigid enough to provide mechanical strength
to the cell. Second, the size of bacterial cells is so small that
transportation of cargo by pure diffusion would be sufficiently rapid
for the survival of the cell. These general considerations and the lack
of direct evidence for cytoskeletal structures in the early experiments 
on prokaryotes led to the common belief that the prokaryotic cells lack 
a cytoskeleton. However, more recent experimental evidences strongly 
indicate the existence of bacterial homologs of the filamentous proteins 
in eukaryotic cells. For example, FtsZ is a bacterial homolog of tubulin 
whereas MreB and ParM are those of actin. Moreover, CreS (crescentin) is 
considered to be a strong candidate for the bacterial counterpart of 
intermediate filaments of eukaryotic animal cells. However, so far it 
has not been possible to identify any bacterial homolog of the eukaryotic 
motor proteins. Nevertheless, the existence of such homologs with very low 
sequence identity with their eukaryotic counterparts cannot be ruled out. 

FtsZ polymerize to form protofilaments. But, unlike eukaryotic tubulins,
these protofilaments do not cooperate to form higher order tube-like
structures which would be analogoues to microtubules. On the other hand, 
ParM polymerizes to form a double-stranded helical filament which is also
very similar to filamentous F-actin.



\subsection{Nucleation of MT and actin filaments} 

The role of $\gamma$-tubulin in the nucleation of MT filaments has been 
known for quite some time. Two classes of actin nucleating proteins are: \\
(i) formin protein family; and 
(ii) Arp2/3 complex.\\



\subsection{Dynamics of polymerization/depolymerization of MT and actin: treadmilling and dynamic instability} 

The dynamics of polymerization and depolymerization of microtubules is 
quite different from those of most of the common proteins. 
{\it Dynamic instability} is now accepted
as the dominant mechanism governing the dynamics of microtubule
polymerization. Each polymerizing microtubule persistently grows for
a prolonged duration and, then makes a sudden transition to a
depolymerizing phase; this phenomenon is known as ``catastrophe''.
However, the rapid shrinking of a depolymerizing microtubule can get
arrested when it makes a sudden reverse transition, called ``rescue'',
to a polymerizing phase. It is now generally believed that the dynamic
instability of a microtubule is triggered by the loss of its
{\it guanosine triphoshate} (GTP) cap because of the hydrolysis of
GTP into {guanosine diphosphate} (GDP). But, the detailed mechanism,
i.e., how the chemical process of cap loss induces mechanical
instability, remains far from clear.

Some of the fundamental quantitative questions on this phenomenon are 
as follows: Does the system reach a steady state under the given 
conditions, and if so, what is the distribution of the lengths of 
the microtubules in that state? It has been discovered that some small 
molecules can suppress the dynamic instability and infuence the  
rates of growth and/or shrinkage of the microbules when bound to 
the tubulins. These molecules are potential anti-cancer drugs because 
of the corresponding implications of the dynamic instability in cell 
division.  What are the quantitative effects of these drug molecules 
on the nature of the steady state (if it still exists) and on the 
corresponding distribution of the microtubule lengths?



\section{Push and pull of cytoskeletal filaments: nano-pistons}

In this section, we focus only on the mechaisms of force generation 
by polymerizing cytoskeletal filaments, namely, microtubules and 
actin. However, these phenomena will be reconsidered again in part IV
in the broader contexts of cell motility and cell division.

A microtubule can keep growing even when it encounters a microsopically 
light obstacle; its action in such situations is reminiscent of a piston. 
Unlike microtubules, the protofilaments of FtsZ do not exhibit dynamic 
instability. On the other hand, unlike actin, filamentous ParM exhibits 
a dynamic instability which is very similar to that of microtubules 
except that the instability of ParM filaments is caused by the hydrolysis 
of ATP rather than that of GTP. Moreover, unlike actin, whose polar
polymer grows aymmetrically through {\it treadmilling}, ParM exhibits
a symmetrical bidirectional growth where rate of elongation at both ends
are identical.



\subsection{Spring-like force generated by cytoskeletal filaments}

The piston-like action of polymerizing filaments is not the only mode 
of motor-independent force generation. Spring-like actions of 
filamentous structures are known to drive fast motility of some 
biological systems. One well known example of such biological spring 
is the vorticellid spasmoneme whose major protein component is spasmin. 
The sperm cell of the horse-shoe crab {\it Limulus polyphemus} also 
utilizes the spring-like action of a coiled bundle, which consists 
mainly of actin filaments, to penetrate into an egg for its fertilization.



\section{Processive cytoskeletal motors: porters}

Many cytoskeletal motors carry molecular cargo over distances which are 
quite long on the intracellular scale. Because of their superficial 
similarities with porters who carry load on their heads, these motors 
are often colloquially referred to as ``porters''.



\subsection{Architectural designs of the porters: common features}

All the cytoskeletal motor proteins have a {\it head} domain;  this domain
contains a site for ATP hydrolysis and another site for binding to a
cytoskeletal filament which serves as a track for the motor. Binding of
ATP to the head alters the affinity of the motor for its track. The head
domain of the kinesins is the smallest (about 350 amino acids), that of
myosins is of intermediate size (about 800 amino acids) whereas the head
of dyneins is very large (more than 4000 amino acids).

The ``identity card'' for members of a superfamily is the sequence
of amino acids in the motor domain. The members of a given superfamily 
exhibit a very high level of ``sequence homology'' in their motor domain. 
But, the amino acid sequence as well as the size, etc. of the other 
domains differ widely from one member to another of the same superfamily.
All kinesin and dyneins have a {\it tail} domain which binds with the 
cargo. The tail domain exhibits much more diversity than the head domain 
because of the necessity that the same motor should be able to recognize 
(and pick up) wide varieties of cargoes.

Myosins are actin-based motor proteins. According to the widely accepted
nomenclature, myosins are classified into families bearing numerical 
(roman) suffixes (I, II, ..., etc.). According to the latest standardized 
nomenclature of kinesins, the name of each family begins with the word 
``kinesin'' followed by an arabic number ($1$, $2$, etc.). Moreover, 
large subfamilies are assigned an additional letter ($A$, $B$, etc.) 
appended to the familty name. For example, kinesin-14A and kinesin-14B 
refer to two distinct subfamilies both of which belong to the family 
kinesin-14.

Dyneins are microtubule-based motor proteins. Dyneins can be broadly
divided into {\it two} major classes: (i) cytoplasmic dynein, and
(ii) axonemal dynein. Structural features of these motors is quite 
different from those of kinesins and myosins.



\subsection{Mechano-chemistry of cytoskeletal motors: general aspects}

Even for a given single motor domain, a large number of chemical states
are involved in each enzymatic cycle. In principle, there are, many
{\it pathways} for the hydrolysis of ATP, i.e., there are several
different sequences of states that defines a complete hydrolysis cycle.
Although, all these pathways are allowed, some paths are more likely
than others. The most likely path is identified as the {\it hydrolysis
cycle}.



\subsection{Fundamental questions}

We phrase the questions in such a way that these may appear to be 
directly relevant only for the cytoskeletal motors. But, these
can be easily rephrased for the other types of motors including, for
example, those which move on nucleic acid strands. These questions
are as follows:\\
(i) {\it Fuel}: What is the {\it fuel} that supplies the (free-)energy 
input for the motor? The free energy released by the hydrolysis of ATP 
is usually the input for cytoskeletal motors. \\
(ii) {\it Engine, cycle and transmission}: The site on the motor, 
where ATP is hydrolyzed, can be identified as its engine. 
What are the distinct states of the cyclic engine in
various stages of each cycle? Which step of the cycle is responsible
for the generation of force (or, torque)? How is the structural
(conformational) change, caused by this force (or torque), amplified
by the architecture of the motor? In other words, how does the
{\it trasmission} system of the motor work, i.e., what are the analogues
of the {\it clutch and gear} of automobiles? \\
(iii) {\it Track and traction}: Are they filamentous tracks static 
or dynamic, i.e., do the lengths and/or orientations of the tracks change 
with time? What is the traction mechanism used by a motor head for 
staying on track? \\

(iv) {\it Number of engines and coordination of their cycles}:
The state of oligomerization of the motor subunits has important
functional implications. Majority of the members of myosin and 
kinesin superfamilies are homodimers although monomeric, hetero-dimeric 
and tetrameric kinesins have also been discovered. Some members of 
myosin and kinesin superfamilies are known to self-assemble into 
higher-order structures; the most well known among these higher-order 
structures is the myosin thick filaments in muscles which will be 
described later in the context of muscle contraction. What functional 
advantages arise from oligomerization? Are the cycles of the different 
engines of a motor coordinated in any manner and, if so, how is this 
coordination maintained? \\
(v) {\it Stroke and step sizes}: The separation between the two 
successive binding sites on the track is the smallest possible step 
size of the motor. On the other hand, a stroke is a conformational 
change of the motor bound to the track. In general, the stroke size
need not be equal to the step size. What is the stroke size of a 
givn motor? If the motor covers only a fraction of the distance to 
the next binding site by the stroke, how does it manage to cover the 
remaining distance?  Can the same motor adopt different step sizes 
under different circumstances? \\
(vi) {\it Directionality and processivity}: 
Majority of myosins are + end directed i.e., move towards the barbed
end of actin filaments. Similarly, majority of the kinesins are also
+ end directed motors whereas most of the dyneins are - end directed
motor proteins. What determines the direction of movement, i.e., why 
are some motors $+$-end directed whereas the others are $-$-end directed? 
Can a motor reverse its direction of motion (a) spontaneously, or 
(b) under an opposing (load) force?  Do the motors possess reverse 
gears and is it possible to reverse the direction of their movement 
by utilizing the reverse gear mechanism? What is the minimal change 
(e.g., mutation) required to reverse the direction of motion of a 
motor?

One of the key features of the dynamics of cytoskeletal motors is their
ability to attach to and detach from the corresponding track. A motor
is said to be attached to a track if at least one of its heads remains
bound to one of the equispaced motor-binding sites on the correspoding
track. Moreover, a motor can detach completely from its track.

One can define processivity in three different ways:\\
(i) Average number of {\it chemical cycles} in between attachment and
the next detachment from the filament;\\
(ii) {\it attachment lifetime}, i.e., the average time in between an
attachment and the next detachment of the motor from the filament;\\
(iii) {\it mean distance} spanned by the motor on the filament in
a single run.\\
The first definition is intrinsic to the process arising from the
{\it mechano-chemical} coupling. But, it is extremely difficult to
measure experimentally. The other two quantities, on the other
hand, are accesible to experimental measurements.

To translocate processively, a motor may utilize one of the two
following strategies:\\
strategy I: the motor may have more than one track-binding domain
(oligomeric structure can give rise to such a possibility quite
naturally). Most of the cytoskeletal motors like conventional two-headed
kinesin use such a strategy. One of the track-binding sites remains
bound to the track while the other searches for its next binding site. \\
strategy II: it can use a ``clamp-like'' device to remain attached to
the track; opening of the clamp will be required before the motor
etaches from the track. Many motors utilize this strategy for moving
along the corresponding nucleic acid tracks.
The {\it duty ratio} is defined as the average fraction of the time 
that each head spends remaining attached to its track during one cycle.
The typical duty ratios of kinesins and cytoplasmic dynein are at least 
$1/2$ whereas that of conventional myosin can be as small as $0.01$.
What is the mechanism that decides the {\it processivity} (or the lack 
of processivity) and the duty ratio of a motor? \\

(vii) {\it Stepping pattern of a double-headed motor}:
Does the motor move like an ``inchworm'' or does the stepping appear
more like a ``hand-over-hand'' mechanism? Moreover, two types of
hand-over-hand mechanism are possible: symmetric and asymmetric.
In the symmetric pattern, the two heads exchange positions, but the
three-dimensional structure of the molecule is preserved at all equivalent
positions in the cycle. In contrast, in the asymmetric pattern, the two
heads exchange position, but alternate steps differ in some way, e.g.,
what happens in ``limping'' which involves alternate faster and slower
stepping phases. Can a motor switch from one track to a neighbouring
track and, if so, how does it achieve that? What prevents a motor from
changing lane on a multi-lane track? \\
(viii) {\it Speed and efficiency}: Is the average speed of a processive
motor determined by the track or the motor or fuel or some external
control mechanism? Recall that the average speed of a car on a highway
in sparse traffic can be decided either by the smoothness of the highway,
or by the model of the car (whether it is a Ferrari or a heavy truck),
or by the quality of the fuel. Similarly, how does the molecular
constitution of the track and the nature of the motor-track interaction
affect the speed of the motor?  Is the mechano-chemical coupling
{\it tight} or {\it loose}? If hydrolysis of ATP provides the input
free energy, then, how many steps does the motor take for every molecule
of ATP hydrolyzed, or, equivalently, how many ATP molecules are consumed
per step of the motor?  What is the maximum speed it can attain?  Can an 
external force applied to a motor in the forward direction speed it up? 
How does the speed of the motor depend on the opposing ``load'' force? 
As the load force increases, the velocity of the motor decreases. The 
magnitude of the load force at which the average velocity of the motor 
vanishes, is called the {\it stall force}. What happens when the load 
force is increased beyond the stall force? Three possible scenarios are 
as follows: (i) the motor may detach from the track, or (ii) the motor 
may reverse its direction of motion (and move in the direction of the 
load force) (a) without hydrolyzing ATP, or (b) hydrolyzing ATP. 
The force-velocity relation is one of the most fundamental characteristic 
property of a motor. What is the most appropriate definition of
efficiency of the motor and how to estimate that efficiency? \\
(ix) {\it Regulation and control}: How is the operation of the motor
{\it regulated}? For example, how is the motor switched on and off?
Recall that the speed of a car can also be regulated by imposing the
some speed limit or by traffic signals. Are there molecular {\it
signals} that control the motor's movement on its track and how?
How does the motor pick up its cargo and how does it drop it at the
target location? How do motors get back to their starting points of the 
processive run after delivering their cargo?\\
(x) {\it Motor-motor interactions}: How do different types of motors
interact while moving on the same track carrying their cargo? How do
different classes of motors, which move on different types of tracks,
coordinate their functions and even transfer or exchange their cargoes? \\



\subsection{Motility assays}

There are two geometries used for {\it in-vitro} motility assays: 
(i) the {\it gliding} assay and (ii) the {\it bead}
assay.  In the gliding assay, the motors themselves are fixed to a
substrate and the filaments are observed (under an optical microscope)
as they glide along the motor-coated surface. In the bead assay, the
filaments are fixed to a substrate. Small plastic or glass beads, whose
diameters are typically of the order of $1 \mu m$, are coated with the
motors. These motors move along the fixed filaments carrying the bead
as their cargo. The movements of the beads are recorded optically.



\subsection{Myosin porters}



\subsection{Kinesin porters}

$\bullet$ {\bf Homodimeric conventional kinesin: members of Kinesin-1 family}  

Kinesin-1, the prototypical kinesin motor consists of three major
domains:\\
(i) The {\it motor domain}: this domain can be further subdivided
into the globular catalytic core, the adjacent neck linker and the
neck region. The core motor domain consists of about 325 amino acids.\\
(ii) The {\it stalk} which is a $\alpha$-helical coiled-coil domain.\\
(iii) The globular {\it tail} domain at the end of the stalk which can
bind with cargo.



$\bullet$ {\bf Heterodimeric kinesins: members of Kinesin-2 family} 



$\bullet$ {\bf Monomeric kinesins: members of Kinesin-3 family}



$\bullet$ {\bf Kinesin-13 and kinesin-8 familes: MT polymerases and depolymerases}  

Some kinesin motors move diffusively to target one of the two ends of 
the MT track and, then, start depolymerizing the track itself; MCAK, 
a member of the kinesin-13 family, is an example of such kinesins. 
The diffusive motion of the MCAK does not require ATP, but it hydrolyzes 
ATP to power its depolymerase activity. These depolymerases play 
important roles in some crucial stages of cell division which we'll 
take up in the last part of this resource letter.
 
Kip3p, a member of the kinesin-8 family, is also a MT depolymerase. 
But, unlike, MCAK, it ``walks'', rather than diffusing, in a specific 
direction on the MT track by hydrolyzing ATP and, after reaching the 
target end, starts depolymezing the MT.

A single depolymerase can peel off more than one MT subunit from the 
tip of the MT; the larger the number of MT subunits it peels off, 
the higher is its {\it processivity}.



\subsection{Dynein porters}


\section{Intracellular porters: collective transport}

In the preceeding section we have considered operational mechanisms of
processive single cytoskeletal motors. Even in multimeric porters, 
different motor domains coordinate in a certain way that leads to the 
processive directed motion of the motor as a whole. In this section, 
we focus on their collective transport properties which arise from their  
coordination, cooperation, competition, etc. ; collective properties 
of rowers will be taken up later.

The situation envisaged here corresponds to a typical bead-assay where 
the filamentary tracks are fixed to a substrate and motors are attached 
to a micron-sized bead (usually made of glass or plastic). The movement 
of the bead in the presence of ATP is monitored using appropriate optical
micropscopic methods. In such situations, each bead is likely to be 
covered by $N$ motors where, in general, $N > 1$. More than one motor 
is also used for transportation of vesicles and large organelles {\it 
in-vivo}.



\subsection{Load sharing on fixed MT track: single cargo hauled by many kinesins}

Normally, a vesicle or an organelles can be hauled by more than one 
motor simultaneously. If all the motors carrying the cargo are 
plus-end directed (or, if all are minus-end directed) they share the 
load. In order to understand the cooperative effects of such multiple 
motors, in-vitro experiments are easier to perform than in-vivo 
experiments. The force-velocity relations for such cargoes can be 
measured using, for simplicity, micron-size dielectric beads, instead 
of vesicles or organelles.



\subsection{Load sharing on fixed MT track: single cargo hauled by many dyneins}

Cooperative effects of multiple dynein motors exhibits richer 
variety of phenomena.



\subsection{Tug-of-war on fixed MT track: bidirectional transport of a single cargo hauled by kinesins and dyneins}

It is well known that some motors reverse the direction
of motion by switching over from one track to another which are oriented
in anti-parallel fashion. In contrast to these types of reversal of
direction of motion, we consider in this section those reversals where
the cargo uses a ``tug-of-war'' between kinesins and dyneins to execute
bidirectional motion on the same MT track. Several possible functional 
advantages of bidirectional transport have been conjectured.

Wide varieties of bidirectional cargoes have already been identified so
far; these include organelles (for example, mitochondria) as well as
secretaory vesicles and even viruses. If motion in one direction dominates
overwhelmingly over the other, it becomes extremely difficult to identify
the movement unambiguously as ``bidirectional'' because of the limitations
of the spatial and temporal resolutions of the existing techniques of
imaging.

The main challenge in this context is to understand the mechanisms
of this bidirectional transport and those which control the duration of
unidirectional movement in between two successive reversals. This insight
will also be utilized for therapeutic strategies. For example, the motor
or the motor-cargo link may be targeted blocking the virus that hijacks
the motor transport system to travel towards the nucleus. On the other
hand, a virus executing bidirectional movements can be turned away from
the outskirts of the nucleus by tilting the balance in favour of the
kinesins.

At least three possible mechanisms of bidirectional transport have been
postulated. (i) One possibility is that either only + end directed motors 
or only - end directed motors are attached to the cargo at any given 
instant of time. Reversal of the direction of movement of the cargo
is observed when the attached motors are replaced by motors of opposite
polarity. (ii) The second possible mechanism is the closest to the
real life ``tug-of-war''; the competion between the motors of opposite
polarity, which are simultaneously attached to the same cargo and tend
to walk on the same filament generates a net displacement in a direction
that is decided by the stronger side. (iii) The third mechanism is
based on the concept of regulation; although motors of opposite polarity
are simultaneously attached to the cargo, only one type of motors are
activated at a time for walking on the track. In this mechanism, the
reversal of the cargo movement is caused by the regulator when it
disengages one type of motor and engages motors of the opposite polarity.
For experimentalists, it is a challenge not only to identify the
regulator, if such a regulator exists, but also to identify the mechanism
used by the regulator to act as a switch for causing the reversal of
cargo movement. Dynactin has been identified as a possible candidate
for the role of such a regulator.



\subsection{Unidirectional traffic of many cargoes on a single track: Molecular motor traffic jam}

Most of the multi-motor phenomena we have considered in the preceeding
section are restricted to sufficiently low densities where direct
interaction of the cargoes did not occur. As the cargoes are always
much bigger than the motors (in-vitro as well as in-vivo), direct
steric interactions of the cargoes become significant when several
cargoes are carried by sufficiently dense population of motors along
the same track. Such situations are reminiscent of vehicular traffic
where mutual hindrance of the vehicles cause traffic jam at sufficiently
high densities. In analogy with vehicular traffic, we shall refer to
the collective movement of molecular motors along a filamentary track
as ``molecular motor traffic''; we shall explore the possibility of
molecular motor traffic jam and its possible functional implications.

Most of the minimal theoretical models of interacting molecular motors
utilize the similarities between molecular motor traffic on MT and 
vehicular traffic on highways both of which can be modelled by 
appropriate extensions of the totally asymmetric simple exclusion 
process. In such models the motor is represented by a ``self-propelled'' 
particle and its dynamics is formulated as an appropriate extension of 
the dynamics of the totally asymmetric simple exclusion process. 
In such models, in addition to forward `hopping'' from one binding 
site to the next, the motor particle is also allowed to detach from 
the track. Moreover, attachment of a motor particle to an empty site 
is also allowed. 

In reality, a molecular motor is an enzyme that hydrolyses ATP and 
its mechanical movement is coupled to its enzymatic cycle. In some 
recent works on cytoskeletal motor traffic, the essential features 
of the enzymatic cycle of the individual motors have been captured.



\subsection{Bidirectional traffic of many cargoes on a single track: Molecular motor traffic jam}



\subsection{Cargoes at crossings and on park-and-ride transport system}

Filamentous actin forms branched networks. Therefore, naturally, a 
fundamental question on transport of cargoes by unconventional 
myosins on actin networks is what happens to the cargo when it 
reaches a point where a single track branched out in two different 
directions. Moreover, similar situation also arises when a cargo 
hauled by kinesins reaches a crossing of MT tracks.

Furthermore, the networks of microtubules and actin filaments are 
not disconnected. The cytoskeleton is microtubule-rich near the cell 
center whereas dense actin filaments dominate the cytoskeleton near 
the cell periphery. On their way to destinations near the cell 
periphery, cargoes cover some distance by taking ride on  
microtubule-based kinesin motors and then switch to actin-based 
myosin motors; this is sometimes referred to as the park-and-ride 
transport system in analogy with that in metro cities of developed 
nations. Similar transfer of cargoes from actin network to 
microtubule network during the transport in the reverse dorection 
is also well documented.



\subsection{Intracellular transport of vesicles and organelles: general aspects}



\subsection{Examples of intracellular transport and traffic: axonal transport}

In a human body, the axon can be as long as a meter whereas the
corresponding cell body is only about 10 microns in length. Almost all
the proteins needed to maintain the synapses are synthesized in the
cell body. How are these proteins transported to the synapse along
the long axon? The problem is even more challenging in animals like
elephant and giraffe which have even longer axons.
A budle of parallel MTs usually run along the axons and dendrites; these
form the track for the motorized transport of vesicles and organelles.
In axons the plus end of the MTS point towards the axonal presynaptic
terminus (the growth cone in developing neurons and the end-plate in
motor neurons), i.e., from the center to the periphery of the cell. In
contrast, the plus ends of the MTs in dendrites are mostly oriented
towards the center of the cell. Movement of the cargo in a direction
away from the cell body is called {\it anterograde} whereas that in
the reverse direction is called {\it retrograde}; the former is driven
by kinesins while the latter is dominated by dyneins.



\subsection{Examples of intracellular transport and traffic: intraflagellar transport}



\subsection{Examples of intracellular transport and traffic: tip growth}

Cytoskelton plays crucial role also in creating and growing tip-like cell 
surface protrusions. Examples of such tip growth phenomena include axonal 
elongation in mammals, growth of root hairs and pollen tubes in plants,
hyphal growth in filamentous fungi, etc.



\subsection{Diseases caused by malfunctiong of cytoskeletal motor transport system}

Just as occasional disruption of work in any department of a factory
can bring entire operation factory to a standstill, defective molecular
machines can cause diseases. For example, malfunctioning of the track
and/or motor can cause breakdown of the intracellular molecular motor
transport system leading to a traffic-jam-like situation



\subsection{Hijacking of cytoskeletal transport system by viruses}

Viruses are known to hijack the motors to travel from the
cell periphery to the cell nucleus. 



\subsection{Drug delivery using cytokeletal motors}

The molecular motor transport system can be utilized even
for targeted drug delivery where molecular motors can be used as
vehicles for the drug.



\section{Processive cytoskeletal motors: cross-linkers and sliders}

\subsection{Cross-linking and relative sliding of two MT filaments by kinesin}

Eg5, a tetrameric kinesin, one of the most prominent members of the 
kinesin-5 family, has been studied experimentally because of it is 
believed to slide two microtubule filaments with respect to each other 
in the mitotic spindle. Although Eg5 is processive, its processivity 
is quite low. Sliding of MT filaments, driven collectively by Eg5, 
has some  similarities with myosin-driven muscle contraction, which 
we'll consider in a later section. 



\subsection{Cross-linking and relative sliding of two actin filaments by processive myosin}

Cross linking and relative sliding of actin filaments by myosin motors 
can, in prinicple, create fingerlike cell protrusions called filopodia.
The sliding of actin filaments by nonprocessive muscle myosins will be 
considered later in the context of muscle contraction. 



\subsection{Axonemal dynein and beating of flagella}

The molecular composition, structure and dynamics of eukaryotic flagella 
are totally different from those of bacterial flagella. Moreover, 
structurally, eukaryotic flagella and cilia are qualitatively similar 
cell appendages; their quantitative differences lie in their size and 
distribution on the cell.

In this subsection we consider only the physical processes driven by 
the cytoskeletal filaments and the motors which lead to the beating 
of the flagella. How the various patterns of these beatings in a 
fluid medium propels the eukaryotic cell is a problem of fluid dynamics 
and will be aken up later in the section on swimming of eukaryotic 
cells.



\section{Nonprocessive cytoskeletal motors: collective dynamics of rowers}

The oars of rowers come in contact with water for a very brief period, 
giving a stroke and then comes out of water, completing one cycle. 
All the rowers of the same group try to synchronize their stroke 
cycle in such a way as to provide the maximum thrust to the boat. 
Similarly, ``rower'' molecular motors also remain attached to their 
track for a small fraction of their ATPase cycle, i.e., the duty 
ratio of these nonprocessive motors is usually small. However, the 
collective stroke of a very large number of such tiny motor molecules 
can generate forces large enough to contract a muscle. 



\subsection{Nonprocessive myosin and muscle contraction}

There are some chemical differences between the muscles of vertebrates
and invertebrates (e.g, flight muscles of insects). Muscle cells of
vertebrates can be broadly classified into ``striated'' and smooth
(``non-striated'') types. Vertebrate striated muscle cells can be
further divided into two categories- skeletal and cardiac. Although
skeletal muscles of vertebrates (e.g., those of frog and rabbit)
were used in most of the early investigations on the mechanism of
muscle contraction, the cardiac muscle has been getting attention in
recent years because of its implications in cardiac disease control. 

Each muscle fiber is actually an enormous multinucleated cell
produced by the fusion of many mononucleated precursor cells
during development whose nuclei are retained in the adult
muscle cell. The diameter of muscle cells is typically $10-100$
$\mu$m and the length can range from less than a millimeter to
a centermeter. Each of these cells is enclosed by a plasma membrane.
The nuclei are squeezed to the peripheral region just beneath
the plasma membrane.

About $80$ percent of the cytoplasm of a skeletal muscle fiber
(i.e., muscle cell) is occupied by cylindrical rods of protein
and are known as myofibrils. Many myofibrils, each about $1 \mu$m
in diameter, are contained within the cross section of a single
muscle cell. The mucle cells also contain mitochondria sandwiched
between the myofibrils.

Myofibrils are the structures that are responsible for muscle
contraction. The most distinctive feature of myofibrils is their
banded appearance; the dark bands correspond to higher density
of protein. The spatial periodicity of the alternating light and
dark bands is 2.3-2.6 $\mu$m in the resting state of a muscle;
the entire repeating structure, from one $Z$-disc to the next,
is known as sarcomere.

The banded appearance of the sarcomere is produced by hundreds of
protein filaments bundled together in a highly ordered fashion.
The two main types of filament are:\\
(i) thick filaments, about 15 $nm$ in diameter, are made mostly
of myosin; \\
(ii) thin filaments, about 7 $nm$ in diameter, consist mostly of
actin.\\
Both these types of filaments contain also other types of proteins
which help to hold them in correct steric arrangement and regulate
the process of contraction.

Arrays of thin and thick filaments overlap in the sarcomere in a
manner similar to that of two stiff bristle brushes.
Myosin molecules are arranged in such a way on the thick filament
that their heads point away from the mid-zone towards either end
of the filament. The thick filaments come within about 13 nm of the
adjacent thin filament which is close enough for the formation of
{\it cross-bridges} between the myosin heads belonging to the thick
filament and actin molecules constituting the thin filaments.

In two landmark papers published in 1954, A.F. Huxley and Niedergerke
and, independently, H. E. Huxley and Hanson
proposed the {\it sliding filament hypothesis} of
muscle contraction. According to this hypothesis, it is the sliding
of the thick and thin filaments past each other, rather than folding
of individual proteins, that leads to the contraction of the muscle.
This theory was formulated clearly and quantitatively in another 
classic paper of A.F. Huxley in 1957.

In the original version of the sliding filament model, developed in the
nineteen fifties, it was generally assumed that the cross bridges moved 
back and forth along the backbone of the thick filaments remaining 
firmly attached to it laterally. However, later X-ray studies 
demonstrated that the filament separation could vary without apparently 
interfering with the actin-myosin interactions. On the basis of this 
observation, in 1969, H.E. Huxley proposed the myosin ``lever arm''  
hypothesis. This model was developed further and formulated quantitatively 
by A.F. Huxley and Simmons in 1971.



Studies of muscle contraction has a long history.



Some of the classic papers on the molecular structure and mechanism 
of muscle contraction is very rewarding even today.



Theoretical models have now reached a high level of sophistication.



\subsection{Bidirectional motion of MT driven collectively by nonprocessive kinesins}

Consider a group of identical motors bound to an elastic backbone. 
Even if each individual motor is non-processive, such a system of
elastically coupled motors can move collectively on a filamentary
track in a processive manner in one direction for a period of time
and, then, spontaneously reverse its direction of motion. Such 
spontaneous oscillations can account for the dynamics of axonemes,
which are core constituents of eucaryotic cilia, as well as 
oscillatory motions of flight muscles of many insects.  


\section{Cooperative extraction of membrane tubes by cytoskeletal motors}

Cytoskeletal motors carry membrane-bounded ovesicles and organelles 
as cargoes while walking along their respective tracks as porters. 
Interestingly, motors can also extract membrane tubes from vesicles. 
The nature of the dynamics of the tube, however, depends on the 
extent of processivity of the motors. 



\section{Effects of defect and disorder on shuttles and muscles}

So far we have implicitly regarded the microtubule track for the 
cytoskeletal motors to be a perfectly periodic array of motor-binding 
sites. However, in reality, the MAPs can introduce ``defect'' and 
``disorder'' into this perfectly periodic lattice; the lattice 
constant being 8 nm. In particular the Tau proteins are known to 
block the kinesin-binding sites on the microtubules. Binding of Tau 
affects at least two different rate constants, namely, those 
corresponding to: (a) the  attachment of a new motor to the track, 
and (b) the forward stepping of the motor.  



\section{Self-organization of microtubule-motor complex in-vitro}

In the earlier sections in this part we have focussed attention on 
situations where motors move on filamentous tracks that neither change 
length nor orientation during the entire period of movement of the 
motors. We have also separately considered the dynamic instability of 
microtubules because of which microrubules can grow or shrink. In this 
section we study the interplay of the dynamics of both microtubules 
and motors, addressing the question of the structures that emerge from 
the self-organization of microtubule-motor complex in-vitro. 
These are also relevant for the phenomenon of cell division which will 
be taken up in the last part of this resource letter.




{\bf Part II: Molecular machines for synthesizing, manipulating and degrading macromolecules of life}

The individual {\it monomeric residues} that form proteins and nucleic
acids are {\it amino acids} and {\it nucleotides}, respectively. 
Both these types of macromolecules are {\it unbranched} polymers.
The {\it complete covalent} structure is called the {\it primary} 
structure of the macromolecule. It would be extremely time- (and space-) 
consuming to write a chemical formula for the entire primary structure. 
Therefore, it is customary to express primary structures in terms of 
abbreviation using an alphabetic code. The most common convention uses 
one-letter code for each nucleotide and three-letter code for each amino 
acid. For proper biological function, these macromolecules form 
appropriate {\it secondary} and {\it tertiary} structures. The term 
{\it conformation} is synonymous with tertiary structure.



In this part we consider molecular machines which either synthesize, or 
manipulate, or degrade macromolecular constituents of a cell, namely, 
DNA, RNA or protein. Among these machines, some translocate along a 
macromolecular track whereas others, whose spatial positions are more 
or less fixed, translocate macromolecules. This difference, of course, 
corresponds to a mere change of reference frame. Many of these machines 
are involved in all the major biological functions of genetic materials, 
e.g., transcription, replication, repair and recomobination as well as 
in defence system of the cell against invading foreign genetic materials.

The constituent monomers, the primary structures as well as the spatial
organization of the higher-order structures of nucleic acids and proteins
are quite different. However, in spite of these differences, there are
some common features in the birth, maturation and death:

(i) The sequence of the monomeric subunits to be used for synthesis
are dictated by the corresponding template. \\
(ii) both nucleic acids and proteins are made from a limited number of
different species of monomeric building blocks. \\
(iii) these polymers are elongated, step-by-step, during their birth by
successive addition of monomers, one at a time. \\
(iv) Synthesis of each chain (polynucleotide and polypeptide) begins
and ends when the machine encounters well-defined start and stop signals
on the template strand.\\
(v) The primary product of the synthesis, namely, polynucleotide or
polypeptide, often requires ``processing'' whereby the modified
product matures into functional nucleic acid or protein, respectively.\\
(vi) DNA, the genetic blueprint of life, needs repair every now and then
to maintain its integrity. During cell division, it is faithfully
replicated and passed onto the daughter cells. But, the other
macromolecules of life are not so lucky. Because of ``wear and tear'',
these aged macromolecule becomes less useful with the passage of time.
Finally, these are degraded, i.e., shredded into its constituent subunit
pieces which are, then, recycled for synthesis of fresh macromolecules.

Interestingly, all these processes are driven by molecular machines 
which nature has designed for the specific purpose.



\section{Packaged organization of nucleic acids}

In every cell the genetic information is encoded in the sequence of
the nucleotides. Thus, at some stage of biological evolution, Nature
chose an effectively {\it linear} device (namely, a NA strand) and a
{\it quaternary} code (i.e., four symbols, namely, A, T, C, G) for
storing genetic information. This was not the most efficient choice!
The fewer is the number of letters of the alphabet the longer is the
string of letters required to express a given message. One serious
consequence of nature's choice of the memory device and coding system
is that even for the most primitive organisms like an {\it E.coli}
bacterium, the total length of the DNA molecule is orders of magnitude
longer than the organism itself! The problem is more acute in case of
eukaryotic cells where an even longer DNA has to be accomodated within
a tiny nucleus! Moreover, random packaging of the DNA into the nucleus
would not be desirable because, for wide variety of biological processes
involving DNA, specific segments of the DNA molecules must be ``unpacked''
and made accessible to the corresponding cellular machineries.
Furthermore, at the end of the operation, the DNA must be re-packed.
Nature has solved the problem of packaging genetic materials in the
nucleus of eukaryotic cells by organising the DNA strands in a
hierarchical manner and the final packaged product is usually referred
to as the {\it chromatin}.



Even in bacteria and viral capsids, the genome has to be packaged in 
a manner which allows efficient access during various processes of 
DNA and RNA metabolism. Most often, packaging or unpackaging and 
repackaging of the genome requires specific molecular machines. 
We'll consider some of these machines later in this part of this 
article.



\section{Elasticity of macromolecules of life} 

Nature must have extracted some advantage from the synthesis of
macromolcules during the course of biological evolution. If it
could manage all biological functions with small molecules, living
systems would not consist with such a large component of macromlecules.
What new features did macromlecules introduce? It not only introduced
a new length scale (characterized by its size) and a time scale
(associated with its dynamics) but also brought in its ``flexibility''
which is not possible with only small molecules. This flexible nature
of macromlecules also gives rise to the importance of conformational
entropy. In fact, many biological processes are driven by entropic
elasticity. Apriori, it is not at all obvious that the phenomenological 
concepts of classical theory of elasticity, which were developed for 
macroscopic objects, should be applicable even for single molecules 
of DNA, RNA, etc. Technological advances over the last two decades 
made it possible to stretch, bend and twist a single macromolecule and 
the corresponding moduli of elasticity have been measured.



Thus, elasticity of macromolecules of life is an interesting topic of 
research in its own right. Moreover, most often genome (DNA or RNA) are 
stored in bent conformation. For example, in eukaryotic cells, DNA is 
bent and wrapped around histones. Similarly, in viral capsids, nucleic
acids are strongly bent for efficient packaging. Furthermore, temporary 
bending of macromolecues take place in many biological processes driven 
by molecular motors. Therefore, the elasticity of the macromolecules 
of life is also interesting in the study of molecular machines which 
polymerize, manipulate and degrade these molecules.



\section{Rings and bracelets} 

A large number of molecular machines, which perform diverse functions 
in DNA metabolism, have toroidal architecture that is a characteristic 
feature of their multi-domain or multi-meric structure. The possible 
functional advantages of the toroidal architecture might have been 
exploited by nature in its evolutionary design of its intracellular 
toolbox.



\subsection{Clamps and clamp loaders}

Not all processive motors have a naturally strong grip on the nucleic 
acid track. Such motors hold onto the track during their translocation 
because they are attached to ring-like special clamp; such clamps are 
loaded onto the track by some other special purpose ATP-consuming 
machines called clamp loaders.



\subsection{SMC proteins}

Members of the SMC ({\underbar{S}}tructural {\underbar{M}}aintenence of 
{\underbar{C}}hromosomes) family of proteins are bracelet-like devices which 
are found in both prokaryotes and eukaryotes. Thse form core components 
of the cohesin and condension complexes in eukaryotes. A common feature 
of their architectural design is the two ``arms'' which are connected 
at a hinge. ATP-driven conformational transformations of these machines 
manifest as transitions from ``opening'' of the two arms of a bracelet  
about the hinge which joins them. 

Now we focus exclusively on the ATP-dependent operational mechanism of 
the SMC proteins. Later, we'll consider their role in important processes 
of DNA metabolism, e.g., in chromosome segregation.



\section{Helicase and unzipping of nucleic acids}

Helicases are molecular motors that unzip double-stranded nucleic acids 
and translocate along one of the two strands. Some Helicases also 
function as ``sweepers'' in the sense that non-helicase proteins bound 
to the nucleic acid strand are dislodged by a helicase. 

Nucleic acid translocases either move along nucleic acid tracks or, 
if anchored, move a nucleic acid strand. Helicases are special types 
of nucleic acid transocase as these translocate along single strands 
of nucleic acids by unzipping double-stranded nucleic acids. 
There are many nucleic acid translocases which, in spite of structural 
similarity with helicases, do not unzip nucleic acids. 

Now we focus only on the mechanisms of operation of helicases. Later 
we'll examine their operational mechanisms in broader contexts like, 
for example, replication, repair and recombination. 

Helicases have been classified in various ways using different criteria.

(i) Several conserved amino-acid sequences have been discovered in
helicases. On the basis of these ``helicase signature motifs'', 
DNA helicases have been classified into superfamilies SF1, SF2, SF3, 
etc.\\
(ii) On the basis of the nature of the nucleic acid (DNA or RNA) track,
i.e., the nucleic acid which they unwind, helicases have been classified
into (a) DNA-helicases, (b) RNA-helicases and (c) hybrid helicases.
Some helicases are, however, hybrid in the sense that these can 
unwind both DNA and RNA. \\
(iii) Some helicases move from 3' to 5' end of a ssDNA whereas others 
move in the opposite direction. On the basis of directionality, 
helicases have been classified into two groups: 3' to 5' helicases 
and 5' to 3' helicases. \\
(iv) Helicases have also been grouped according to the the source of
these proteins, i.e., humans, plants, bacteria, viruses, etc. \\
In this chapter, we study the mechanisms of helicases separtely for
monomeric, dimeric and hexameric helicases.
(v) On the basis of the number of ATPase domains, helicases have been
classified into monomeric and multimeric types; dimeric and hexameric
being the most common multimeric helicases.\\
Here we utilize this last scheme of classification for highlighting 
the unity of mechanisms of helicases in spite of their diverse functions.

One of the fundamental questions on the mechano-chemistry of helicases 
is the mechanism of energy transduction- does it unzip the nucleic acid 
actively or does it utilize the transient opening of the double strands 
by thermal fluctuations in a manner which can be identified as a thermal 
ratchet mechanism?



Helicases are required for unwinding double-stranded DNA molecules.



Some other helicases unwind RNA molecules. However, few helicases 
are capable of unwinding both DNA and RNA molecules.



A few helicases are monomeric. Very generic models of helicase motors 
have been developed, which may be interpreted as theories of monomeric 
helicases. 



Dimeric helicases are more common; the stepping pattern of these are 
expected to be analogous to double-headed conventional cytoskeletal 
motors. Analog of the hand-over-hand mechanisms of the cytoskeletal 
motors is called the ``rolling'' model. However, most of the dimeric 
helicases are believed to follow the inchworm mechanism.



A large number of helicases are hexameric and have an approximate 
ring-like architecture. For hexameric helicases, at least three 
alternative mechanisms of enzymatic activities have been suggested;
these include, activities of all the ATP-binding domains in
(i) parallel, (ii) random, (iii) sequential manner. \\

(i) Parallel: In this mechanism all the subunits hydrolyze dTTP and
exert power stroke simultaneously. \\
(ii) Random: there are at least two possible different scenarios: \\
(a) random in time, where power stroke of each subunits starts and
finishes at random times independent of other units;
(b) random in space, where power strokes are sequential in time (i.e.,
each subunit can begin only after another finishes), but the order of
power strokes around the ring is random.\\
(iii) Sequential: there are at least two different sequences in which
the subunits can exert power stroke:\\
(a) paired sequential, i.e., sequentially around the ring, but with
diametrically opposite subunits in the same state;
(b) ordered sequential, i.e., sequential in the strict order 1,2,...6
around the ring.



\section{Topoisomerases and untangling of nucleic acids} 

During various processes in DNA metabolism, often DNA strands get 
entangled. Topoisomerases untangle nucleic acids thereby changing 
their topology. The extent of supercoiling is expressed quantitatively 
by the {\it linking number} which is the sum of the {\it twist} 
and {\it writhe} of the DNA molecule. The linking number is an integer 
and is a topological characteristic property of the molecule. DNA 
molecules with different linking numbers are called topoisomers. 
The topoisomerase interconverts topoisomers and hence the name. 

Topoisomerases are divided into two classes which are named type I 
and type II. Type I topoisomerases can change the linking number 
of a closed circular DNA in steps of $\pm 1$ whereas type II 
topoisomerases change the linking number in steps of $\pm 2$. 
This is achieved by type I topoisomerases by first cleaving one strand 
of the DNA and, then, after passing the other strand through this 
break, resealing the break. In contrast, a type II topoisomerase 
cleaves both strands of a dsDNA and passes another intact segment of 
dsDNA through this break. Type I and II topoisomerases are further 
classified into subfamilies designated as IA, IB, IIA, IIB, etc. on 
the basis of primary sequence and operational mechanism. DNA gyrase 
of {\it E-coli} are among the most extensively studied topoisomerases. 
Reverse gyrase, as the name suggests, introduces supercoiling opposite 
to that introduced by the gyrase. Besides positive and negative 
supercoiling, the two other types of reactions catalyzed by 
topoisomerases are (i) knotting or unknotting, and (ii) catenation or 
decatenation. 

The mechanisms of type I topoisomerases are simpler that those of 
type II topoisomerases. Each of the type II topoisomerase machines 
consist of two identical halves and two ``gates''. The untangling 
of DNA occurs through sequential opening and closing of these gates 
appropriately coordinated with a transient nicking and subsequent 
ligation of one of the two strands which are thus made to pass 
through each other.



\section{Membrane-associated machines for macromolecule translocation: exporters and importers} 

We now consider the translocation of three types of macromolecules, 
namely, DNA, RNA and proteins across cell membranes as well as internal 
membranes of eukaryotic cells. In the next part, we'll consider 
active transport of small molecules and ions across membranes by 
molecular machines.

\subsection{Export and import of macromolecule across membranes: general principles} 

Macromolecules to be translocated across the pore may be
{\it hydrophobic} or may be electrically charged. Therefore, it is
not surprising if it encounters an energy barrier while trying to
translocate across the pore. However, what makes macromolecule
translocation even more interesting from statistical physics
perspective is that the macromolecule also encounters an
{\it entropic} barrier. The number of allowed conformations of the
macromolecular chain, and hence its entropy, is drastically reduced
when it translocates across a narrow pore. Therefore, in general,
the barrier encountered by the translocating macromolecular chain
is a {\it free energy} barrier.  

So far as the process of macromolecule translocation is concerned,
it can be divided into two sub-processes- in the first, the tip of
the macromolecule just enters the pore and, then, in the second
subprocess the entire length of the chain crosses the pore. The
first process is analogous to putting the tip of a thread through
the hole of a needle whereas the second is the analogue of pulling
a length $L$ of that thread through the same hole after successful
insertion of the tip. Both power stroke and Brownian ratchet mechanisms 
have been proposed. Power stroke can manifest itself either as a 
``push'' or a ``pull'' in the appropriate direction. 



\subsection{Export and import of DNA} 

$\bullet$ {\bf DNA translocation through nanopores: general principles} 



$\bullet$ {\bf DNA transfer across cell membranes: viral and bacterial DNA} 

Three basic mechanisms of intercellular DNA transfer in bacteria are:\\
(i) {\it Transformation}, i.e., uptake of naked DNA (DNA which is
not associated with proteins or other cells) from extracellular
environment;\\
(ii) {\it Transduction}, i.e., indirect transfer of bacterial DNA
into a new cell by a bacteriophage;\\
(iii) {\it Conjugation}, i.e., direct transfer of DNA between two
bacteria which are in physical contact with each other.



$\bullet$ {\bf DNA transport through eukaryotic nuclear pore complex} 

The nuclear pore complex (NPC) is itself a large aseembly of proteins;
the inividual protein components of this assembly are called
{\it nucleoporins}. This assembly has a {\it eight-fold} symmetry
about an axis normal to the plane of the membrane. On the cytoplasmic
(i.e., exterior) side of the membrane, {\it eight fibrils} extend
from the eight lobes which are arranged in the form of a ring. On
the nucleoplasmic (i.e., interior) side of the membrane these eight
fibers join to form a {\it basket-like} structure at a distance of
approximately $100$ nm from the inner membrane. DNA uptake into the 
nucleus through the nuclear pore complex are received little attention 
so far.



\subsection{mRNA export from eukaryotic nucleus} 

The m-RNA must be exported from the nucleus before it can be translated 
into proteins. 



\subsection{Export and import of proteins} 

$\bullet$ {\bf Protein translocation across membranes: general principles} 

Protein translocation can take place (a) during synthesis
({\it co-translation}, e.g., in ER), or (b) after completion of
synthesis ({\it post-translation}, e.g., in mitochondria).



$\bullet$ {\bf Bacterial protein secretion machineries} 



$\bullet$ {\bf Machines for protein import across nuclear membrane in eukaryotes} 

Proteins are synthesized in the cytoplasm. Those proteins which function 
inside the nucleus must be imported from the cytoplasm. 



$\bullet$ {\bf Machines for protein translocation across membranes of organelles} 

There are two distinct major pathways of protein transport in eukaryotic 
cells: (i) the vesicular pathway, and (ii) non-vesiular pathway.  In 
the vesicular pathway, proteins are transported from one membrane-bounded 
organelle to another after packing the protein in a vesicle. The vesicle 
buds out from the donor organelle and, after reaching the destination, 
fuses with the acceptor organelle. In this pathway, the vesicle is 
transported in the cytoplasmic environment by cytoskeletal motor 
transport system which we have discussed earlier in part I of this article. 



Next we focus on the non-vesicular pathway where proteins are translocated 
across membranes of organelles by protein-translocating machines. 
 


$\bullet$ {\bf Machines for protein translocation across membranes of endoplasmic reticulum} 



$\bullet$ {\bf Machines for protein translocation across membranes of mitochondria and chloroplasts} 

Most of the proteins are translocated into mitochondria post-translationally.
Mitochondria have a translocase of the outer membrane (called TOM) and a 
translocase of the inner membrane (called TIM). Similarly, the corresponding 
translocases of chloroplasts are names as TOC and TIC, respectively.
The twin-arginine translocation (Tat) pathway of the thylakoid membrane of 
chloroplasts and their prokaryotic counterparts share some common features.



$\bullet$ {\bf Machines for protein translocation across membranes of peroxisome} 



\section{Genome packaging machines of viral capsids} 

As stated earlier, the viral genomes may consist of DNA or RNA. There 
are two alternative mechanisms for packaging of the genome. In case 
of some viruses, the genome is encapsulated by molecules that  
self-assemble around it. In contrast, the genome of other viruses are 
packaged into a pre-fabricated empty container, called {\it viral 
capsid}, by a powerful motor. As the capsid gets filled, The pressure 
inside the capsid increases which opposes further filling. The effective 
force, which opposes packaging, gets contributions from three sources: 
(a) bending of stiff DNA molecule inside the capsid; 
(b) strong electrostatic repulsion between the negatively charged  
strands of the DNA; 
(c) loss of entropy caused by the packaging. 

One of the model systems, which has been very popular among the researchers, 
is the bacteriophage $\phi29$; its genome consists of a double-stranded DNA. 
For understanding the mechanism of packaging double-stranded RNA into the 
viral capsids, the bacteriophage $\phi6$ has been used as model system.



The highest pressure generated inside the capsid of the $\phi29$ is about 
$60$ times the normal atmospheric pressure (i.e., about $10$ times 
the pressure in a typical champagne bottle!) and the corresponding force 
applied by the packaging motor is about $60$ pN. Thus, genome packaging 
motors of viral capsids are among the strongest discovered so far. 
What is the mechanism used by these motors to generate such a relatively 
large force (large compared to the forces generated by most of the other 
motors)?

At first sight, the phenomenon seems to have (at least superficially) 
several similarities with translocation of macromolecules into 
eukaryotic organelles. Therefore, questions on the mechanisms of 
translocation motors can also be reformulated for understanding the 
mechanisms of packaging motors of viral capsids. Are the nucleic acids 
``pulled'' or ``pushed'' into the capsid head by the motor? Or, is the 
mechanism better described by a ``nut-like'' rotation of the packaging 
motor on the ``bolt-like'' nucleic acid strand?



\subsection{Energetics of packaged genome in capsids}



\subsection{Structure and mechanism of viral genome packaging motor}



\section{Polynucleotide polymerases} 

Among the macromolecules of life, nucleic acids and proteins are 
polymerized by machines which use the respective tracks also as 
templates for the synthesis. In this section, we consider polymerase 
machines which synthesize nucleic acids while translocating on another 
nucleic acid. The main quantities of interest in this context is 
the rate of synthesis of the macromolecules. Although most of the 
works initially focussed on the average rates, the fluctuations in 
the rate of synthesis is receving more attention in recent years 
because of two recent developments: (a) the availability of 
experimental techniques for detection of individual macromolecular 
products as they are synthesized and released, and (b) the relevance 
of transcriptional and translational noise in the study of overall 
noise in gene expression.

The free energy released by the polymerization of the polynucleotide 
products serve as the input energy for the driving the mechanical 
movements of the corresponding polymerase. Therefore, these are also 
regarded as molecular motors. 
Polymerase motors generate forces which are about 3 to 6 times stronger 
than that generated by cytoskeletal motors. But, the step size of a 
polymerase is about 0.34 nm whereas that of a kinesin is about 8 nm. 
Moreover, the polymerase motors are slower than the cytoskeletal motors 
by two orders of magnitude. Furthermore, natural nucleic acid tracks are 
intrinsically inhomogeneous because of the inhomogeneity of nucleotide 
sequences whereas, in the absence of MAPs and ARPs, the cytoskeletal 
tracks are homogeneous and exhibit perfect periodic order. 

The polymerase is expected to have binding sites for (a) the template 
strand, (b) the nascent polynucleotide strand, and (c) the NTP subunits.
It must have a mechanism to select the appropriate NTP dictated by 
the template and a mechanism to catalyze the addition of the NTP 
thus selected to the growing polynucleotide. It must be able to step 
forward by one nucleotide on its template without completely 
destabilizing the iternary complex consisting of the polymerase, the 
template and the product. Finally, it must have mechanisms for 
initiation and termination of the polymerization process for which, 
most often, it requires assistance of other proteins. 

Most of the fundamental questions we raised in the context of the
cytoskeletal motors remain valid also for polynuclotide polymerases.
Some further questions, that are unique for polymerases, are posed
below:\\

(i) Are the two translocations, namely nucleotide addition and
forward movement of the polymerase, tightly coupled? Or, is it
possible to add nucleotide to the growing product without forward
movement of the polymerase? In principle, the latter seems to be
possible provided the conformation of the TEC changes accordingly.

(ii) What are the paths of the template and product polynucleotide
chains within the polymerase? If the template one of the two strands
of a double-stranded nucleic acid, what path does the non-template
strand follow?

(iii) Does the template and the nascent product polynuclotide form
any hybrid structure and, if so, what are the (free-)energetics of
the that determine the maximum size of the hybrid? What causes the
product polynucleotide to separate from the corresponding template?

(iv) Do the secondary structures of the template and the product
play any role in the process of polymeraization?



On the basis of the nature of the template and product polynucleotides, 
polymerases can be broadly divided into four classes: 
DNA-dependent RNA polymerase (DdRP), DNA-dependent DNA polymerase (DdDP), 
RNA-dependent DNA polymerase (RdDP) and RNA-dependent RNA polymerase 
(RdRP).

There are several common architectural features of all polynucleotide
polymerases. The shape of the polymerase has some resemblance with the
``cupped right hand'' of a normal human being; the three major domains
of it are identified with ``fingers'', ``palm'' and ``thumb''. There
are, of course, some crucial differences in the details of the
architectural designs of these machines which are essential for their
specific funstions. The most obvious functional commonality between
these machines is that these add nucleotides, the monomeric subunits
of the nucleic acids, one by one following the template encoded in the
sequence of the nucleotides of the template. However, in spite of the 
gross architectural similarities between the polymerases in prokaryotic 
and eukaryotic cells, there are significant differences in the 
primary sequences of these machines.

The main stages in the synthesis of polynucleotides by the polymerase
machines are common:

(a) {\it initiation}: Once the polymerase encounters a specific sequence
    on the template that acts as a chemically coded start signal, it
    initiates the synthesis of the product. This stage is completed when
    the nascent product becomes long enough to stabilize the macromolecular
    machine complex against dissociation from the template. \\
(b) {\it elongation}: During this stage, the nascent product gets
    elongated by the addition of nucleotides.  \\
(c) {\it termination}: Normally, the process of synthesis is terminated,
    and the newly polymerized full length product molecule is released,
    when the polymerase encounters the {\it terminator} (or, stop)
    sequence on the template. However, we shall consider, almost
    exclusively, the process of {\it elongation}.

\subsection{DdRP and transcription} 

In all kingdoms of life, the DdRP are multi-subunit enzymes. The 
eukaryotic DdRP machines are not only larger in size than their 
bacterial counterparts, but also consist of larger number of subunits.
There are three different types of DdRP in eukaryotic cells, namely,
RNAP-I, RNAP-II and RNAP-III. The mRNA, which serves as the template
for protein synthesis, is polymerized by RNAP-II whereas rRNA and
tRNA are synthesized by RNAP-I and RNAP-III, respectively. 

A common architectural feature of all DdRPs is the ``main internal
channel'' which can accomodate of DNA/RNA hybrid that is typically
$8$ to $9$ bp long. The NTP monomers enter through another pore-like
``entry channel'' while the nascent transcript emerges through the
``exit channel''. The formation of the bond between the newly arrived
NTP and the RNA chain takes place at a catalytically active site
located at the junction of the entry pore and the main channel.
In principle, during actual transcription, it may be necessary first
to unwind the DNA, at least locally, to get access to the nucleotide
sequence on a single-stranded DNA. Interestingly, the RNAP itself
exhibits helicase activity for this purpose.



The translocation of a polymerase along its template resembles that
of a device that moves along a digital tape and reads information
from it. The synthesis of the product polymerase is, then, analogous
to writing of new information. However, unlike digital logic of a
computer, decisions made by a polymerase are governed by competing
rates and equilibria among alternative conformations and complexes.
The decisions which regulate its operation are dictated by two types
of input: intrinsic and extrinsic. Discrete segments of the template
and product polynucleotides, with which the polymerase interacts,
provide intrinsic inputs. Extrinsic inputs come from small ligands
and other regulatory proteins.

Single molecule studies of DdRP have provided quantitative data on 
the force-velocity relation for these motors.



But, use of these techniques have also led to the discoveries of new 
phenomena. For example, in the initiation stage, the DdRP can 
{\it scrunch}.



During the elongation stage, the DdRP can not only ``pause'' for unusually 
long time, but can also ``backtrack'' which can, occasionally, lead 
to premature termination of transcription, unless the machine re-starts 
its forward movement.



Most recent progress in single-molecule imaging has led to the 
discovery of ``transcriptional bursts''.



Quantitative modeling of the DdRP in transcription began almost two 
decades ago. The collective movement of DdRP on a given track is 
interersting from several different perspectives. For example, a 
stalled DdRP can be restarted by another approaching it from behind 
and such a scenario can lead to polymerization of the transcripts in 
a ``burst''.





\subsection{Primase: a unique DdRP } 

DdDP cannot begin polymerization of a polynucleotide from scratch. 
First, a DNA primase polymerizes a short RNA primer using the DNA 
template. Then, a DdDP adds nucleotide subunits to the primer thereby 
continuing DNA replication. One of the fundammental questions is 
how does the primase coordinate its operation with those of the 
DdDP machines?



\subsection{DdDP and DNA replication} 

Two DdDP machines have to replicate the two complementary strands of 
DNA both of which serve as templates. However, each DdDP translocates  
unidirectionally ($5' \to 3'$) elongating the product strand. As a 
result, the ``leading strand'' is synthesized processively, whereas 
the ``lagging strand'' is replicated discontinuously; the ``Okazaki 
fragments'' synthesized by this discontinuous process are then joined 
together (ligated). The coordination of the operation of the two 
polymerases is one of the interesting aspects of the operational 
mechanism of the DdDP machines.



Single molecule manipulation of DdDP have elucidated the operational 
mechanisms of these machines and resulted in the recent progress in 
their quantitative modeling.



The DdDP alone cannot replicate the genome; together with DNA clamp and 
clamp loader, DNA helicase and primase, it forms a large multi-component  
complex machinery which replicates the DNA and is often referred to as 
the {\it replisome}. 



Transcription of a gene is carried out a large of times during the life 
time of a single cell. In contrast, a distinct feature of DNA replication 
is that, during its lifetime, a cell must not replicate its genome more 
than once. Only recent investigations have explored how cell achieves 
this requirement. 



\subsection{RdDP and reverse transcription} 

A reverse transcriptase is a RdDP which uses a RNA template to polymerize 
a DNA. The most common example of RdDP is the HIV-1 reverse transcriptase 
which synthesizes DNA from the RNA genome of the human immunodeficiency 
virus (HIV). HIV-1 reverse transcriptase is one of key targets for the 
some of the drugs which are being tried against AIDS.



\subsection{Telomerase: an unique RdDP}

Telomeres, i.e., telomeric DNA, are the terminal DNA at chromosome ends. 
Telomerase is a unique reverse transcriptase that uses an RNA template 
to polymerize telomeric DNA. In the absence of telomerase operation, 
telomerers would gradually shorten in each round of DNA replication 
because the DdDP cannot replicate these end portions of the DNA. 
Shortening of tolemere is believed to be a cause of premature ageing 
and other age-related diseases. Therefore, understanding the operational 
mechanism of telomerase will help in the control of premature ageing as 
well as in developing cancer therapeutics. 



\subsection{RdRP and RNA replication} 

In spite of strong resemblance of the overall shape of all the RDRPs
with a ``cupped right hand'', viral RDRPs have some special architectural
features. The most notable distinct feature of
these polymerases is that, in contrast to the ``open hand'' shape of
the other polynucleotide polymerases, the RDRP resembles a ``closed
hand''. The closing of the ``hand'' is achived by loops, called
``fingertips'', which protrude from the fingers and connect with the
thumb domain at their other end. The fingertip region forms the
entrance of the channel where the RDRP binds with the RNA template.
In addition, there is a small positively charged tunnel through which
the nucleotide monomers required for elomgation of the RNA enter.
The genome of some of the viruses consist of double stranded RNA;
the corresponding RDRP have some additional unique structural elements
which unzip the two strands and feed the appropriate strand to the
catalytic site.



\subsection{Nucleic-acid analogs as templates for polynucleotide polymerase} 

Nucleic acid analogs with altered backbones or bases have been synthesized 
artificially. Threose nucleic acid (TNA) is a nucleic acid analog whose 
backbone consists of repeating threose units linked by phosphodiester 
bonds. Glycerol nucleic acid (GNA) is another analog of natural nucleic 
acids. GNA is based on glycerol-phosphate backbone repeat unit. 
The operational mechanism of polynucleotide polymerases using these 
nucleic acid analogs as templates has become a subject of experimental 
investigation in recent years. Understanding these processes may shed 
new light on the origin of life as some of the nucleic acid analogs 
might have been used the genetic mterial by the earliest forms of life 
on earth.



\subsection{Coordination between transcription and replication} 



\subsection{Transcription and replication of DNA in mitochondria and chloroplast} 

Mitochondrial DNA (mtDNA) is replicated by a mtDdDP called DNApol $\gamma$.
However, the mechanism of replicating the lagging strand of mtDNA is 
different from that of replicating the nuclear DNA. 

Surprisingly, in majority of the eukaryotes, the mitochondrial DdRP is 
structurally closer to single-subunit polymerases of bacteriophages, rather 
than the multi-subunit polymerases of bacteria, although mtDNA are believed 
to have bacterial ancestor. 



\section{Ribosomes and polymerization of polypeptides} 

Synthesis of each protein from the corresponding messenger RNA (mRNA)
template is carried out by a ribosome and the process is referred to 
as {\it translation} (of genetic code). Ribosome is one of the largest 
and most sophisticated macromlecular machines within the cell. Even 
in the simplest organisms like single-cell bacteria, a ribosome is 
composed of few rRNA molecules as well as several varieties of protein 
molecules.

Each ribosome consists of two parts which are usually referred to as
the large and the small subunits. The small subunits binds with the 
mRNA track and assists in decoding the genetic message encoded by the 
codons (triplets of nucleotides) on the mRNA. But, the actual polymerization 
of the protein (a polypeptide) takes place in the large subunit. The 
operations of these two subunits are coordinated by a class of adapter 
molecules called tRNA.

The ``head'' and the ``body'' are the two major parts of the {\it small} 
subunit. Two major lobes, which sprout upward from the ``body'', are 
called the ``platform'' and the ``shoulder'', respectively. The decoding 
center of the ribosome lies in the cleft between the ``platform'' and 
the ``head'' of the small subunit. The incoming template mRNA utilizes a 
``channel'' formed between the ``head'' and the ``shoulder'' as a conduit 
for its entry into the ribosome. Through the cleft between the ``head'' 
and the ``platform'' the mRNA exits the ribosome.

The characteristic ``crown-like'' architecture of the {\it large} 
subunit arises from three protuberances. On the flat side of the large 
subunit exists a ``canyon'' that runs across the width of the subunit 
and is bordered by a ``ridge''. Halfway across this ridge, a hole leads 
into a ``tunnel'' from the bottom of the ``canyon''. This ``tunnel'' 
penetrates the large subunit and opens into the solvent on the other 
side of the large subunit. This ``tunnel'' serves as the conduit for 
the exit of the nascent polypeptide chain. This ``tunnel'' is 
approximately 10 nm long and its average width is about 1.5 nm.

Several intersubunit ``bridges'' connect the two subunits of each
ribosome. This bridges are sufficiently flexible so that relative
movements of the two subunits can take place in each cycle of the
ribosome. The intersubunit space is large enough to accomodate just 
three tRNA molecules which can bind, at a time, with the three 
binding sites E, P and A. Moreover, the shape of intersubunit space 
is such that it allows easy passage of the L-shaped tRNA molecules.

Just like the synthesis of polynucleotides (e.g., transcription and 
replication), synthesis of polypeptides (i.e., translation) also 
goes through three stages, namely, {\it initiation}, {\it elongation}, 
and {\it termination}. During the elongation stage, the three major 
steps in the chemo-mechanical cycle of a ribosome are as follows: 
In the first, the ribosome selects a aa-tRNA whose anticodon is exactly 
complementary to the codon on the mRNA. Next, it catalyzes the reaction 
responsible for the formation of the peptide bond between the existing 
polypeptide and the newly recruited amino acid resulting in the elongation 
of the polypeptide. Finally, it completes the mechano-chemical cycle by
translocating itself completely to the next codon and is ready to begin
the next cycle. 

Elongation factors (EF), which are themselves proteins, play important 
roles in the control of these major steps which require proper 
communication and coordination between the two subunits. The need for 
coordination between the two subunits can be appreciated from the 
following considerations. The formation of the peptide bond between the 
growing polypeptide and the newly arriving amino acid (which can take 
place only in the larger subunit) can be allowed only after it is 
recognized as the correct species implied by the genetic code. During 
the process of checking its identity through the condon-anticodon 
matching (which takes place in the smaller subunit), the formation of 
the peptide bond is prevented by an elongation factor Tu (EF-Tu). 
However, once a cognate tRNA is identified, the smaller subunit sends 
a ``green signal'' (by a molecular mechanism that remains unclear), the 
EF-Tu separates out by a process driven by GTP hydrolysis thereby 
clearing the way for the peptide bond formation. Similarly, elongation 
factor G (EF-G) coordinates the translocation of the mRNA by one codon 
and the simultaneous movement of the tRNA molecules from one binding site
to the next one.



\subsection{Single-molecule experiments to probe single ribosome mechano-chemistry}

Thus, each ribosome has three different functions which it performs on
each run along the mRNA track:\\
(i) it is a {\it decoding device} in the sense that it ``reads'' the
sequence of codons on the mRNA and selects a aa-tRNA whose anticodon
is exactly complementary to the codon on the mRNA.\\
(ii) it is a {\it peptidyltransferase} that catalyzes the reaction
responsible for the formation of the peptide bond between the existing
polypeptide and the newly recruited amino acid resulting in the
elongation of the polypeptide.\\
(iii) it is a {\it conveying machine} that, while moving along a mRNA
chain, passes tRNA molecules through itself during elongation.\\
Interestingly, function (i) is performed exclusively by the smaller
subunit while the function (ii) is carried out in the larger subunit.
This division of labour between the larger and the smaller subunit
may be related to the fact that there is a relatively large (8 nm)
separation between the anti-codon and the amino-acid-carrying end of
the tRNA molecules. But, the function (iii) requires coordinated
movement of the two subunits.

Some specific steps in the mechano-chemical cycle of a ribosome are
driven by the hydrolysis of guanosine triphosphate (GTP) to guanosine
diphosphate (GDP). Therefore, ribosome is often regarded as a motor.
However, a ribosome is not merely a ``protein-making motor protein'' but 
it serves as a ``mobile workshop'' which provides a platform where a 
coordinated action of many tools take place for the selection of the 
appropriate subunits and for linking them to synthesize each of the 
proteins. As this mobile workshop moves along the ``assembly line'' 
(mRNA), new subunits (amino-acids) are brought to it by the ``workers'' 
(tRNA molecules).

Some of the fundamental questions on the mechanism of translation are 
the following: \\
(i) How does the tRNA move on the ribosome (a) before, and (b) after 
the peptide bond formation? (ii) How does the ribosome modulate the 
stability of its binding with the mRNA so that it can step forward on 
its track once in each cycle during the elongation stage without 
destabilizing the ribosome-mRNA-tRNA complex itself? (iii) How is the 
movement of the ribosome on mRNA coordinated with the movements of 
the tRNA molecules on the ribosome? (iv) What are the sources of energy 
required for these movements and how are these energies transduced?
Most of these questions can be addressed using single-ribosome techniques; 
however, very few such experiments have been reported in the literature.



Theoretical modeling of single-ribosome operation has made very little 
progress so far. 



\subsection{Traffic-like collective movement of ribosomes}

Most often many ribosomes move simultaneously on a single mRNA strand 
while each synthesizes a separate copy of the same protein. Such a 
collective movement of the ribosomes on a single mRNA strand has 
superficial similarities with vehicular traffic and is, therefore,  
referred to as ribosome traffic. Most of the theoretical models of 
ribosome traffic represent the mRNA as a one-dimensional lattice, 
where each site corresponds to a single codon. Since an individual 
ribosome is much larger than a single codon, the ribosomes are 
represented by hard rods in these models. So, ribosome traffic is  
treated as a problem of non-equilibrium statistical mechanics of a 
system of interacting ``self-driven'' hard rods on a one-dimensional 
lattice. Moreover, in these models the inter-ribosome interactions 
are captured through hard-core mutual exclusion principle: none of 
codons can be covered simultaneously by more than one ribosome. Thus, 
these models of ribosome traffic are essentially totally asymmetric 
simple exclusion process for hard rods: a ribosome hops forward, by 
one codon, with probability $q$ per unit time, if an only if the hop 
does not lead to any violation of the mutual exclusion principle.



But, strictly speaking, a ribosome is neither a particle nor a hard
rod. Moreover, in all the works mentioned above, the entire complexity 
of the mechano-chemistry of each ribosome is captured by a single 
parameter $q$. Only a few attempts have been made in recent years to 
capture the mechano-chemistry of individual ribosomes in the 
quantitative models of interacting ribosomes in trafic-like situations.



\subsection{Multi-machine coordination in gene expression}

Coordination of various machines is a key feature of assembly line in 
manufacturing industries. However, the mechanisms of coordination of 
the machineries of transcription, translation and other steps of gene 
expression are not well understood in spite of investigations in this 
direction.


\subsection{Non-ribosomal machineries of polypeptide synthesis} 

Machines for synthesis of polyketides and non-ribosomal polypeptides 
have been known for long time. But, hardly any quantitative modeling 
of these machines have been made so far.



\section{Spliceosome; closely related to ribosome?}

In eukaryotic cells, pre-mRNA require several types of processing 
before it matures to a functional mRNA. One of the key processes 
is splicing whereby non-coding segments (introns) are removed and 
the resulting strands are joined. Just like the ribosome, spliceosome 
is also composed of proteins as well as RNA.



\section{Fidelity of template-dictated polymerization} 

Nature's evolutionary design has successfully optimized two competing
demands: accuracy and speed of template-dictated synthesis of nucleic
acids and proteins. The typical proability of the errors is about $1$ 
(i) in $10^{3}$ polymerized amino acids, in case of protein synthesis, 
(ii) in $10^4$ polymerized nucleotides in case of mRNA syntheis and 
(iii) in $10^9$ polymerized nucleotides in case of replication of DNA. 
The mechanisms of proof
reading and quality control also has to optimize two other mutually
conflicting demands: maintaining the integrity of the genome and
tolerance for some errors (genetic mutations) which is necessary for
diversification of species. What are the mechanisms used by the
intracellular machinery for manufacturing macromolcules of life to
simultaneously achieve these conflicting goals?



\subsection{Fidelity of DNA replication: polymerization by DdDP}


\subsection{Fidelity of transcription: polymerization by DdRP}


\subsection{Fidelity of polymerization by RdRP}



\subsection{Fidelity of polymerization by RdDP}



\subsection{Fidelity of translation}


\subsection{Connection between fidelity of transcription, translation and replication}


\section{Machines for non-template-dictated biopolymerization}

\subsection{Machines for polymerization of polysachharides}



$\bullet$ {\bf Starch synthesizer machines} 

Most well studied examples of polysaccharide polymerization include 
biosynthesis of starch.



$\bullet$ {\bf Starch synthesizer machines} 

Biosynthesis of cellulose have also received lot of attention.



$\bullet$ {\bf Chitin synthesizer machines}

Machines for synthesizing chitin have been studied for quite some time.



\subsection{Machines for polymerization of fatty acids}



\section{Machines for assembling and remodeling of chromatin} 

As we wrote earlier, in a living cell, DNA double strands are not 
available in naked form. In eukaryotic cells DNA is packaged in a 
hierarchical structure called chromatin. 



There are some common features, in spite of wide range of differences, 
in the organization of eukaryotic and prokaryotic chromosomes.



In order to get access to the relevant segments of DNA for various 
processes in DNA metabolism, eukaryotic cells use a class of  machines 
which alter the DNA-histone interactions. These machines fall in 
two different classes: 
(i) enzymes that covalently modify histone proteins (histone modifying 
enzymes), and (ii) ATP-dependent chromatin-remodeling enzymes (CRE) 
which alter the structure and/or position of the nucleosomes.



\subsection{Histone modifying enzymes} 

In addition to modification of the histones in the nucleosomes, the 
linker histone also plays important role as the ``gate keeper'' by 
regulating the access to nucleosomal DNA.


\subsection{Roles of CRE in chromatin assembly} 

Although, at first sight, chromatin-assembly and remodeling may appear 
to be opposite processes, there is a step common to both- sliding of 
the nucleosome on a DNA. Therefore, it should not be surprising that 
several CRE participate in both these processes.



\subsection{ATP-dependent chromatin-remodeling enzymes} 

Some of the CRE repress chromatin, instead of activating it. Therefore, 
a general definition of chromatin remodeling should be as follows:  
chromatin remodeling is a change in the state of chromatin that 
facilitates its activation or repression. 



$\bullet$ {\bf RSC: a chromatin-remodeling enzyme} 

RSC ({\underline{R}}emodelling the {\underline{s}}tructure of 
{\underline{c}}hromatin) is a well studied remodeller of chromatin 
in yeast.



$\bullet$ {\bf NURF: a chromatin-remodeling enzyme} 

NURF ({\underline{Nu}}cleosome {\underline{r}}emodeling  
{\underline{f}}actor) is a well studied remodeller of chromatin 
in drosophila.


$\bullet$ {\bf CHRAC: a chromatin-remodeling enzyme} 

CHRAC ({\underline{Chr}}omatin {\underline{a}}ccessibility   
{\underline{c}}omplex) is a well studied remodeller of chromatin 
in drosophila.


\subsection{Machines for packaging and manipulating bacterial DNA}

Although bacterial chromosomal DNA are not enclosed in any nucleus, 
these are, nevertheless, packaged in a structure call {\it nucleoid}.



The competing forces, which arise from DNA-protein interactions and 
determine the compact organization of the bacterial DNA have been 
investigated.



\subsection{Machines for packaging and manipulating mitochondral DNA}



\subsection{Chromatin remodeling in transcription; transcription factory} 



\subsection{Chromatin remodeling in replication} 



\section{Machines for DNA Repair and recombination} 

For hereditary transmission of the genome, repair of damaged DNA is 
as essential as the high fidelity of the replication of the genome 
itself. Recombination can be viewed as a process whose sole purpose 
is to rearrange genetic material thereby generating genetic diversity. 
However, recombination can also be used to repair damaged DNA. 
Recombinational repair is not the only method of DNA repair; there 
are other methods of DNA repair which do not exploit recombination.
Just as DNA replication requires coordinated operation of several 
machines, DNA repair and recombination also needs similar coorporation 
of another set of machines.



\subsection{Motor-driven junction migration} 



\subsection{RecA motor in recombination} 



\subsection{RecBCD motor in recombination} 



\subsection{RecG motor in recombination} 



\subsection{RecQ motor in recombination} 



\subsection{RuvABC motor in recombination} 



\subsection{Rad51 and Rad54 motor in recombination} 



\subsection{Mfd motor in repair} 



\section{Machines for degrading macromolecules of life} 

\subsection{Machines for degrading DNA} 

Nucleases are enzymes which function as ``scissors'' by cleaving the 
phosphodiester bonds on nucleic acid molcules. Endonucleases cleave 
the phosphodiester bond within the nucleic acid thereby cutting it i
into two strands whereas exonucleases remove the terminal nucleotide 
either at the 3' end or at the 5' end.



Erythrocytes (red blood cells) and lens fiber cells in the eyes, posses 
no nucleus! In reality, DNA is removed from the precursors of these 
cells during their maturation.  Moreover, during the development of an 
animal, some cells are deliberately killed; this phenomenon, known as 
apoptosis (programmed cell death), also involves degradation of the DNA 
of the target cells. Furthermore, those cells which become toxic or 
senescent are also killed actively and their DNA are degraded. Finally, 
bacteria have evolved a mechanism of degrading DNA of invading 
bacteriophages. In this subsection we list references of some relevant 
papers on the molecular architecture and mechanisms of operation of 
the machines which degrade DNA. 



\subsection{DNA degradation by restriction-modification enzymes} 

R-M systems consist of two components which perform two competing
functions. Restriction involves an endonuclease that breaks a DNA
by hydrolyzing the phosphodiester bond in backbone of both the
strands. On the other hand, modification involves a methyltransferase
which adds a chmical group to a DNA base at a position that blocks
the restriction activity. Both these activities are specific for
the same dNA sequence. The main biological function of the R-M system
is to defend the bacterial host against bacteriophage infection by
cleaving the phage genome while the DNA of the host are not cleaved. 

The restriction endonucleases have been classfied into three groups: 
type I, type II and type III. Both type I and type III are molecular 
machines in the true sense because these require ATP for their 
operation.



\subsection{Machines for degrading RNA} 

Ribonucleases (whose commonly used abbreviation is RNase) are also 
nucleases and function as ``scissors'' that cleave the phosphodiester 
bonds on RNA molecules. Like all other nucleases, RNases are also 
broadly classified into endoribonucleases and exoribonucleases.



$\bullet$ {\bf Exosome: an RNA deagrading machine} 

In eukaryotes, a barrel-shaped multi-protein complex, called exosome, 
degrades RNA molecules. The bacterial counterpart of exosome is usually 
referred to as the RNA degradosome. The fundamental questions on the 
operational mechanism of these machines are of two types. The first 
types of questions are essentially identical to those raised earlier 
in the context of import/export of macromolecules by translocation 
motors. The second type of questions are similar to those raised in 
the context of (ribo-)nucleases, namely, the mechanism of shredding 
or mincing and the resulting size distribution of the products.



\subsection{Machines for degrading proteins} 

Proteases are enzymes which perform functions that are analogous to 
nucleases. Just as nucleases cleave the phosphodiester bonds on 
nucleic acids (i.e., polynucleotides), proteases cleave peptide bonds 
on polypeptides and, hence, sometimes also called peptidase.

Simple proteases in the extracellular space, e.g., the pancreatic 
proteases, digest proteins derived from diets. However, such non-specific 
proteases are not expected to operate in the intracellular space 
because they would indiscreminately cleave all the essential and 
non-defective proteins to their amino acid subunits thereby destroying 
the cell itself. Evolution has designed intracellular machines for 
protein degradation which mince only the unwanted proteins in very 
specialized chambers whose gates open to allow only for only such 
unwanted proteins. Moreover, since mitochondria and plastids had 
bacterial ancestors, it is not surprising to find very similar 
proteases in these compartments.



$\bullet$ {\bf Bacterial proteases}  

In bacteria, few different families of proteases have been found 
which, however, share some common structural and functional features.





$\bullet$ {\bf Proteasome: a protein deagrading machine} 

Proteasome is a large and complex machine for protein degradation. 
It has structural and functional similarities with exosome; what 
exosome does for RNA, proteasome does for proteins. Onbiously, the 
fundamental questions to be addressed are very similar to those 
in the case of exosomes.



\subsection{Machines for degrading polysachharides} 



$\bullet$ {\bf Machines for degrading starch}  



$\bullet$ {\bf Machines for degrading cellulose} 



$\bullet$ {\bf Machines for degrading chitin} 



\section{Mechanisms for searching target sequence on NA}

In most of the processes we have discussed so far a protein or a 
macromolecular complex has to bind a specific site on a DNA. For 
example, in order to initiate transcription, the transcription 
factor must bind with a specific site on the DNA. Similarly, 
sequence-specific binding is required for the operation of 
restriction enzymes. How does the machine target the specific 
site? Is search through an effectively one-dimensional diffusion 
sufficiently rapid? Or, does the search become more efficient 
by a combination of the one-dimensional diffusion with other 
processes?



\section{Effects of inhomogeneities, defects and disorder}

The sequence of nucleotides on naturally occurring nucleic acid strands 
are intrinsically inhomogeneous. Numerical calculations for the given 
inhomogeneous sequence of any specific nucleic acid strand is not very 
difficult. But, for the simplicity of analytical calculations, two 
extreme idealizations are sometimes considered: in one of these the 
actual sequence is replaced by a hypothetical homogeneous sequence 
whereas in the other the sequence is assumed to be completely random.



{\bf Part III: Membrane associated ion transporters and related machines}

A wide variety of machines are associated with either the plasma membrane 
of the cell or with the internal membranes that enclose various organelles 
like, for example, mitochondria. In part II we have considered machines 
which translocate macromolecules across cell memebranes. Now, in this part, 
we focus on machines which transport small and medium size molecules and 
ions across membranes. 

These ``transporters'' can be broadly divided into two categories-
active and passive. {\it Channels} are passive trasporters because these 
allow the passage of molecules or ions down their electro-chemical 
gradients and do not require input energy for performing this task. 
In contrast, active transporters drive molecules or ions against their 
electro-chemical gradients by utilizing some input energy directly or
indirectly. 

The active transporters can be further divided into two classes- primary 
and secondary active transporters.
Primary active transporters include (a) ATP-binding cassette (ABC)
transporters, (b) ion pumping P-type ATPases. 
Primary active transporters use light or chemical energy as input to 
transport molecules and/or ions across a membrane. In fact, one
of the major roles of pumps is to create and maintain electrochemical
gradients by actively transporting ions. Secondary active transporters 
use the spontaneous flow of the ions along such electro-chemical gradients 
to drive other species of molecules ``uphill'' (i.e., against their 
natural own electro-chemical gradients). Interestingly, in spite of 
their mode of operation (i.e., active versus passive), ion channels 
and pumps share one common feature, namely, the ability to transport 
ions in a {\it selective} manner; ion-selectivity is crucial for the 
survival of th cell.



\section{ATP-binding cassette (ABC) transporters: two-cylinder engines of cellular cleaning pumps} 

An ATP-binding cassette (ABC) transporter is a membrane-bound
machine. These machines are found in all cells from bacteria to 
humans. In prokaryotic cells, ABC transporters are located in 
the plasma membrane. In eukaryotes, ABC transporters have been 
found in the internal membranes of organelles like mitochodria, 
peroxisomes, golgi and endoplasmic reticulum. These translocate 
ions, nutrients like sugars and amino acids, drug molecules, 
bile acids, steroids, phospholipids, small peptides as well as 
full length proteins. 

In spite of wide variations in their functions and substrates 
translocated by them, they share some common features of structure 
and dynamics. Each ABC transporter consists of four core domains.
Out of this four, two transmembrane domains (TMDs) are needed
for binding the ligands which are to be transported while the
two nucleotide-binding domains (NBDs) bind, and hydrolyze, ATP.
Many ABC transporters are single four-domain proteins. In contrast, 
``half-size'' ABC transporters conist of one TMD and one NBD; 
many ABC transporters are actually homo-dimers or hetero-dimers 
of ``half-size'' transporters. 

Some of the fundamental questions specifically
related to the mechanisms of ABC transporters are as follows:

(i) why do these machines need two ATP-binding domains although
it consumes only one molecule of ATP for transporting one ligand?

(ii) Do the two NBDs act in alternating fashion, like a
two-cylinder engine where the cycles of the two cylinders are
coupled to each other? Or, do the two NBDs together form a
single ATP-switch?



\section{Membrane associated ion-pumps: P-type ATPases} 

P-type ATPases form a superfamily of machines which transport {\it 
cations} across membranes.



\subsection{Na/K pumps} 

This pump plays crucial roles and is mainly responsible for maintaining
electrolyte balance in almost all cells in humans. It takes in K$^{+}$
ions and ejects out Na$^{+}$ ions.



\subsection{Ca- pumps} 

Cytosolic Ca$^{2+}$ concentration is maintained below about 10 $\mu$M
for normal metabolism of the cell. On the other hand, Ca$^{2+}$ is
one of the most important carriers of signals. During signaling,
brief opening of Ca$^{2+}$ channels in the plasma membrane (or organnellar
membrane) allow Ca$^{2+}$ to enter spontaneously because of the existing
eletro-chemical gradient. However, this increase of Ca$^{2+}$ inside is
only transient as Ca$^{2+}$ pump ejects the Ca$^{2+}$  pump ejects the
Ca$^{2+}$ ions out. It is this Ca$^{2+}$ pump that maintains the high
electro-chemical gradient (low Ca$^{2+}$ concentration inside and
high Ca$^{2+}$ concentration outside).



\subsection{Proton pumps} 

In the fungal plasma membrane (e.g., that of yeast) a proton pump
hydrolyzes ATP to pump out the protons thereby creating an electro-chemical
gradient. This electro-chemical gradient is utilized to provide the energy
to the proton-coupled co-transporters for sugars, amino acids and other
nutrients. The gastric H$^{+}$/K$^{+}$-aTPase is most closely related to 
Na$^{+}$/K$^{+}$ pump; it pumps out H$^{+}$ ions and takes in K$^{+}$ ions.



\subsection{Copper pumps} 

Malfunctioning of the copper-transporting ATPases can lead to Menkes 
and Wilson diseases in humans. 



\subsection{Comparison of P-type ATPases} 



\subsection{Bacteriorhodopsin and halorhodopsin: light-driven proton pumps} 

The input energy for both bacteriorhodopsin (BR) and halorhodopsin (HR) 
is light. But, BR pumps protons whereas HR pumps chloride ions. The 
analog of mechano-chemical cycle of motors is the photocycle of BR and 
HR.



$\bullet$ {\bf Bacteriorhodopsin} 



$\bullet$ {\bf Halorhodopsin} 



$\bullet$ {\bf Xanthorhodopsin} 

Like bacteriorhodopsin, xanthorhodopsin is also a light-driven proton 
pump. But, has a special feature in its structure which, like antenna, 
enables it to collect light more efficiently than what is possible by 
a bacteriorhodopsion.



\subsection{Secondary transporters} 

A secondary transporter can be a symporter or an antiporter. The 
transport of a substrate is coupled to that of either proton 
(utilizing the protonmotive force) or sodium (utilizing the 
sodium-motive force).



\subsection{Membrane-bound ATPases: Some fundamental questions and generic models} 



\section{ATP synthase and related machines} 

ATP synthase is the smallest rotary motor and is embedded in the membrane 
of the organalles mitochondria (in animal cells) and chloroplasts (in 
plant cells). It consists of two coupled parts which are called 
F$_0$ and F$_1$ and, therefore, ATP synthase is also referred to as 
F$_0$F$_1$-ATPase. This motor in reversible. In the normal model, 
F$_0$ is rotated by a protonmotive torque which, in turn rotated 
F$_1$ during which the latter synthesizes ATP from ADP and phosphate. 
In the reverse mode, F$_1$ consumes ATP for its own rotation in the 
reverse direction thereby rotating also F$_0$ in reverse while the 
latter operates effectively as a proton pump.

There are also some interesting architectural similarities between 
the ATP synthase and the TrwB DNA translocase.

\subsection{ATP synthase in mitochondria} 



\subsection{From nutrient to ATP: energy production by mitochondria} 

For the synthesis of ATP by mitochondria, protons must translocate 
across F$_{0}$ subunit of the ATP synthases, which are bound to the 
mitochondrial membranes, down their electro-chemical gradient. But, 
how is the concentration gradient of protons created? This is 
achieved primarily by a process called oxidative phosphorylation 
during which electrons, derived from nutrients, are passed through 
a sequence of enzyme complexes located in the mitochondrial membranes.


\subsection{ATP synthase of plant chloroplasts} 




\subsection{From light to ATP: energy production by chloroplasts} 

The concentration gradient of protons required for ATP synthesis by 
chloroplasts in plants, by harventing sunlight, is the result of a 
sequence of procsses, the primary one being photophosphorylation.



\subsection{Na$^{+}$-ATP synthase} 

We have already seen that Na$^{+}$ can substitute for H$^{+}$ as the 
coupling ion in secondary transporters. Now we point out that even 
for the operation of ATP synthases, H$^{+}$ is not essential and in 
some ATP synthases Na$^{+}$ used instead of H$^{+}$.  


\subsection{Vacuolar ATPase} 


Vacuolar ATPases were initially identified in plant and fungal vecuoles
and hence the name. Later these were found also in plasma membrane and
organelle membranes of mammalian cells and plants. Therefore, it is
more appropriate to link the letter ``V'' in V-ATPase with
``{\it various}'' (various membranes) rather than ``vacuoles''. V-ATPases
are ATP-dependent proton pumps that regulate pH (acidify) intracellular
compartments in eukaryotic cells.



\subsection{Pyrophosphatase} 

Membrane-bound pyrophosphatase (PPase) usually couples pyrophosphate 
(PP$_{i}$) hydrolysis to ion translocation across the membrane.

$\bullet$ {\bf H$^+$-pyrophosphatase} 



$\bullet$ {\bf Na$^+$-pyrophosphatase} 

However, very recently, it has been argued that some bacteria, which 
live under extreme conditions, translocate sodium ion, instead of 
proton, by coupling it to the hydrolysis of pyrophosphate.  





\subsection{Historical notes on ATP synthase and related machines} 




\section{Bacterial flagellar motor} 

In this section, we consider only the rotary motor that drives the 
bacterial flagella. The architecture of the full flagellum and how 
its movements propels the bacterium in the fluid medium will be 
taken up in the section on swimming of bacteria.

\subsection{Bacterial flagellar motor driven by proton-motive force} 

Normally, the bacterial flagellar motor is driven by a protonmotive force, 
i.e., protons driven by a transmembrane electro-chemical gradient. 



\subsection{Bacterial flagellar motors driven by sodium-motive force}

However, flagellar motor in some bacteria are driven by the electro-chemical 
gradient of sodium ions.



\subsection{Comparison between ATP synthase and bacterial flagellar motor}

Both the bacterial flagellar motor and ATP synthase are driven by a 
torque arising from a force which is of electro-chemical origin. 
However, there are also crucial differences between these two rotary 
motors.



{\bf Part IV: Machine-driven cellular processes}

Cells exist in wide variations in their sizes, shapes and internal 
structures. For example, among the bacteria, cocci (spherical), 
bacilli (rod-shaped) and spirochetes (spiral) reflect the shapes 
of these unicellular organisms. Among the unicellular eukaryotes, 
protozoa exhibit some of the most complex and exquisite forms. 
The animal cells also exhibit widely different characteristics. 
The linar dimension of a  typical eukaryotic cell is about 10 $\mu$m.
But, a neuron can be as long as a meter. Each skeletal muscle cell
has more than one nucleus whereas a red blood cell has none at all.
Germ cells have only one set of chromosomes whereas all other cells
have two sets. Hair cells of the inner ear act as mechano-sensors 
while rod cells of the retina of the eye are photo-sensors.
However, in spite of such diversities, there is unity in some of 
the common cellular processes. In this part we consider mainly 
cell motility and cell division both of which are also closely 
related to changes in cell shape. 

\section{Machine-driven cell motility} 

{\it ``From whale sperm to sperm whales, locomotion is almost always 
produced by appendages that oscillate or by bodies that undulate, 
pulse, or undergo peristaltic waves''}- M.H. Dickinson et al., Science 
{\bf 288}, 100 (2000).\\

\subsection{Cell motility: some general principles}



\subsection{Motility of prokaryotic cells}

Unicellular microorganisms have developed diverse molecular mechanisms 
of locomotion. The actual mechanism used by a specific type of organism 
depends on the nature of the environment in the natural habitat of the 
organisms. In this section, we consider exclusively the prokaryotes.

If a bacterium lives in a bulk fluid, it's natural mode of motility 
is {\it swimming}. In contrast, is a bacterium lives in a thin fluid 
film close to a solid surface (i.e., in a wet surface), {\it gliding} 
should be its mechanism of movement. Of course, some bacteria may 
be capable of utilizing both these modes of motility.
However, there are subtleties of swimming and gliding that an uninitiated 
reader may not be able to anticipate. Moreover, there are mechanisms of 
motility other than swimming and gliding. 



\subsection{Bacterial shape and shape changes during motility} 

The shapes of bacterial cells depend, at least partly, on the cytoskeleton.

$\bullet$ {\bf Bacterial shape} 



$\bullet$ {\bf Transient and permanent appendages for motility of prokaryotic cells}

The shape changes during motility and division of cells. Therefore, 
it is not surprising that components of the cytoskeleton play crucial 
roles in both these processes. 

Microorganisms exploit the movements of permanent appendages like 
flagella, cilia, pili and fimbriae for their movements. We have already 
considered the rotary motor which drives bacterial flagella. Although 
the motor is driven by either proton motive force of sodium-motive 
force, the numer of flagella and their spatial arrangments on the cell 
vary widely from one bacterial species to another. Many bacteria 
have only one flagellum whereas some species of bacteria possess more 
than one. Perhaps the most unusual is the periplasmic flagella of 
spirochetes. Moreover, some bacterial species possess dual flagellar 
systems which are suitable for movement under different conditions; 
the polar flagellum is used for swimming in bulk fluids whereas the 
lateral flagella for swarming close to solid surfaces.



$\bullet$ {\bf Swimming of bacteria: flagella driven by rotary motor}  

Depending on the species, a bacterium may have a single flagellum,
or one flagellum at each end, or a tuft of flagella at one or
both ends. Each flagellum consists of a {\sl filament}, a {\sl hook} 
and a {\sl basal body}. The filament is helical and is composed of
eleven protein fibrils arranged like the strands of a rope; a fine 
channel ($\sim 70 A^0$ in diameter) runs through the axis of the 
filament. The hook is a hollow, flexible, proteinaceous structure. 
The basal body consists mainly of the flagellar motor. The flagellar 
motor of a bacterium (say, E-coli) is about 50 nm in diameter and 
consists of about 20 different components. The speed of this rotary 
motor could be of the order of 100 Hz. 

A large class of single-cell bacteria ``swim'' in their aqueous 
environment using their flagella which, in turn, are rotated by 
the flagellar motors driven by proton-motive (or sodium-motive) 
force, as we have already explained earlier. 

The Reynolds number, that characterizes the swimming of bacteria 
in aqueous media, is normally very small. Perhaps, one can 
appreciate the situation better by comparing with swimming of a 
human being; a comparable Reynolds number would be realized if 
a human being tried to swim in honey!



$\bullet$ {\bf Swimming of spirochetes: hidden flagella driven by rotary motor} 

Spirochetes have periplasmic flagella (i.e., flagella which are located 
in the periplasmic space between the outer cell membrane and the cell 
wall). Thus, unlike E-coli, spirochetes do not stick their flagella out 
into the fluid outside the cell. But, the flagella of spirochetes are 
also driven by proton-motive force. Rotation of these flagella deform 
the cell body which, conequently, rolls. It is this corscrew-like 
motion of the spirochete that propels it through the external fluid 
medium. However, one counterintuitive consequence of this mechanism of 
the motility of spirochetes is that  spirochetes move faster in 
gel-like media than in water. 



$\bullet$ {\bf Swimming without flagella: linear motor of spiroplasma} 

The mollicutes ({\it Spiroplasma}, {\it Mycoplasma} and {\it 
Acholeplasma}) are the smallest free-living organisms. Their 
structure is unusual in the sense that they do not have cell 
wall and the standard form of prokaryotic flagella. The 
{\it Spiroplasma} are unique among the mollicutes because, as 
the name suggests, their spiral shape can be viewed as a dynamic 
helical membrane tube (of typical radius of about $0.1 \mu$m). 
They maintain their helical structure by the internal cytoskeletal 
filaments.

The motility of the spiroplasma is driven by its contractile 
cytoskeleton. Thus, in contrast to the rotary motors of flagellated 
bacteria, the machinery driving the spiroplasma are linear motors.
Just like spirochetes, spiroplasma move faster in media with 
higher viscosity. But, in contrast to spirochetes, spiroplasma 
move with higher viscosity irrespective of whether or not it is 
gel-like.



$\bullet$ {\bf Gliding of over surfaces: push of linear motors}  

For some bacteria, like {\it Myxococcus xanthus}, hydration of a slime  
secreted by the bacterium through a nozzle, generates the force required 
for their own movement.



Gliding of {\it Mycoplasma mobile} resembles motion of centipedes 
where tiny ``leg-like'' appendages are powered by ATP hydrolysis.



$\bullet$ {\bf Twitching: pull of linear motor type IV pili}  

Some gliding bacteria use type IV pili for their movement. They first 
extend a pilus that adheres to a substrate; then the pilus is retracted 
and the pull of this hook-like structure propels the bacterium forward.



\subsection{Taxis} 

So far we have discussed the machineries used by prokaryotic cells for 
motility. But, how does the cell sense its environment and decide the 
direction of its motion? 

{\it Chemotaxis} refers to the directional movement in response to the
gradient of concentration of a chemical. Substances which attract a 
cell are called {\sl chemoattractant} while those repelling a cell are 
called {\sl chemorepellant}.  Strictly speaking, {\it aerotaxis} is a 
special case of chemotaxis where the motile cells respond to a gradient 
of concentration of the dissolved oxygen.  {\it Mechanotaxis} is cell 
migration controlled by the rigidity of an underlying substrate.
{\it Phototaxis} is the corresponding response to light gradient whereas 
{\it galvanotaxis} is the ability to move in response to electric 
potential gradient. Finally, {\it haptotaxis} is the motility in response 
to gradient in adhesion of ligands.



\subsection{Motility of eukaryotic cells}

Unicellular eukaryotes, like free-living protozoa, move primarily for 
food. In multicellular prokaryotes, cell locomotion is essential in 
development. Moreover, leukocytes move to offer immune response.
Furthermore, fibroblasts, which are normally stationary, move during 
wound healing.

One of the fundamental questions on cell motility is the molecular 
mechanisms involved in the generation of required forces. Broadly 
speaking, three different mechanisms have been postulated and their 
possibility in specific contexts have been explored: (i) Force 
generated by polymerization of cytoskeletal protein filaments 
(actin and microtubules), (ii) Force generated by cytoskeletal motors 
by their interactions with filamentous tracks, and (iii) forces of 
osmotic of hydrostatic origin.

$\bullet$ {\bf Taxis of eukaryotic cells} 

Chemotaxis is not restricted only to prokaryotes. Eukaryotic cells 
are also guided by an appropriate guidance system. Chemotaxis is 
involved in wide varieties of biological processes starting from 
embryogenesis to wound healing and immune response. Several 
different models of eukaryotic chemotaxis have also been proposed.

{\it Chlamydomonas reinhardtii} is a unicellular biflagellate green 
alga. It has been used extensively as a model experimental system for 
investigating swimming of eukaryotic cells. These cells swim towards 
light by beating their flagella. In contrast to phototaxis of 
{\it C. reinhardtii}, the sperm cells of eukaryotes are guided by 
chemotctic signals. Interestingly, the guidance of the axonal growth 
cones is also guided by chemotaxis.



$\bullet${\bf Crawling of eukaryotic cells: dynamic protrusions} 

The crawling of eukaryotic cells involve the formation and movement 
of transient cell protrusions like lamellipodia, filopodia, etc.
A {\it lamellipodium} is a thin sheet-like protrusion whose typical 
thickness varied between $0.1$ and $0.2$ $\mu$m. In contrast, a 
{\it filopodium} is a finger-like structure whose typical diameter 
varies between $0.1$ and $0.3$ $\mu$m. Structurally, there are crucial 
differences between these two types of cell protrusions. Lamillipodia 
are filled with a {\it branched} network of actin filaments whereas 
{\it parallel} bundles of filamentous actin run along the length of 
filopodia. Quite often filopodia protrude from a lamellipodium. 
Therefore, two different models for the formation of the actin-bundles 
of filopodia have been proposed. In the ``convergent elongation model'', 
the filopodial actin filaments are assumed to originate from the 
lamellipodial actin network. But, in the ``de novo filament nucleation 
model'', the filopodial actin filaments are assumed to nucleate 
separately in the filopodia. A common feature of the actin networks in 
lamellipodia and filopodia is that the fast growing (barbed) ends of 
the actin filaments are oriented towards the membrane which gets 
pushed by the piston-like action of the polymerizing actin filaments. 
This piston-like pushing by polymerizing actin is very similar to 
piston-like action of polymerizing microtubules, which we discussed 
earlier, except that actins can form branched structures whereas 
microtubules do not. Other protrusions of the eukaryotic cell include 
pseudopodia, ruffles, microvilli, invadopodia, etc. 



Two different types of models have been developed in the context of 
cell crawling. Some models focus exclusively on the dynamics of the 
cell protrusions. 



Cell protrusions can also lead to the formation of membrane nanotubes, 
some of which lead to permanent connection between different cells. 



$\bullet${\bf Crawling of eukaryotic cells: full cyclic dynamics}

Detailed models capture all the three stages of 
dynamics - (i) formation of a cell protrusion at the 
leading egde, (ii) adhesion of the leading protrusion to the 
underlying substrate, and (iii) contraction of the cell body accompanied 
by detachment of the rear edge of the cell from the substrate.
The contraction of the cell body and the retraction of the rear are 
dominated by the complex dynamics of a visco-elastic active gel. 
Since this aspect of cell locomotion is beyond the current scope of 
this resource letter, we mention only major reviews and a few important 
theoretical papers.



$\bullet${\bf Gliding of eukaryotic cells}



$\bullet$ {\bf Motility of bacterial pathogens driven by actin comets}  

Bacterial pathogen {\it Listeria Monocyte} uses a simplified mechanism 
of motility based on force generation by actin polymerization. 
In this case a comet-like tail of polymerizing actin filaments push 
the pathogen in the host cell. Unlike, cell crawling, which is also 
driven by actin-polymerization, neither adhesion to a slid susbstrate 
nor retraction of the rear of the cell is required.



$\bullet$ {\bf Motility of nematode sperm by actin-like MSP} 

In contrast to other types of sperm cells which swim using flagella 
(and which we'll consider soon), sperm of several nematode species 
crawl. However, unlike most of the crawling cells, these nematode 
sperms do not contain actin. Instead, another protein, called 
major sperm protein (MSP) acts like actin forming dynamic filaments 
which drive the crawling of the nematode sperm.



$\bullet$ {\bf Swimming of eukaryotic cells: beating of eukaryotic flagella}  

Earlier we have already pointed out that the beating of the eukaryotic 
flagella are driven by axonemal dynein motors which move by hydrolyzing 
ATP.  Eukaryotic cells  beat their cilia not only for motility, 
but, in some circumstances, also to move the surrounding medium 
with respect to the cell surface. For example, the cilia on the 
epithelium of the upper respiratory tracts beat to remove the dust 
and other foreign particles.

Now, the main question is: how does a flagellated eukaryotic cell 
exploit the patterns of beating of its flagellum for its swimming? 
This question is addressed by analying the hydrodynamic effects of 
the the different patterns of beating of the flagellum.



$\bullet$ {\bf Motility of flagellate protozoan from termites}  



\section{Machine-driven cell division} 

\subsection{Cell cycle} 

During its lifetime, before complete division of a parent cell into 
its two daughters, a cell goes through a sequence of states which 
are identified primarily by its shape and internal architecture.



\subsection{Brief introduction to cell division: eukaryotes versus prokaryotes} 

Although each stage of the cell cycle is of interest to cell biologists, 
we are mainly interested in the machines and mechanisms involved in 
the different stages of cell division. In particular we focus attention 
on mitosis and cytokinesis.  



\subsection{Mitosis and chromosome segregation in eukaryotes: machines and mechanisms}

{\it Mitosis} is a complex process whereby identical copies of the 
replicated genome are segregated so as to form the separate genomes 
of the two daughter cells which would result from the cell division. 
The bipolar machinery which carries out this process is called the 
mitotic spindle. A similar machinery, called the {\it meiotic 
spindle}, runs the related process of {\it meiosis}, which reduces 
the size of the genome by half to produce a haploid gamete from a 
diploid one. We shall consider separately a few important sub-steps 
of mitosis. 
 
A large number of coordinated processes are involved in mitosis. 
These include, for example, spindle morphogenesis, chromosome 
condensation, sister-chromatid separation, dynamic instability of the 
microtubules, depolymerase-driven length control of microtubules, 
walking of MT-associated motors on their tracks, etc. We have studied 
several of these active processes separately in the preceeding 
sections. It is the integration of so many processes within a single 
theory of mitosis that poses the main conceptual challenge to 
theoretical modelers.

There are three different sources of forces which govern the dynamics 
of the mitotic spindle: (i) Forces generated by cytoskeletal motors 
which can capture microtubules and can also slide microtubules with 
respect to each other; (ii) pushing and pulling forces exerted by 
polymerizing and depolymerizing microtubules; (iii) spring-like 
forces which arise from the elastic stretching of the chromosomes.
Moreover, the bending of the microtubules may have important consequences.

$\bullet$ {\bf Spindle morphogenesis} 

Both the mitotic spindle and the meiotic spindle are formed by 
microtubules (MT), MAPS and cytoskeletal motors. In principle, 
a spindle can form by one of the two different  pathways. In the 
centrosome-directed pathway, the spindle starts from centrosomes 
(which are eventually located at the poles of the spindle) and grow 
towards the center by adding tubulin subunits at their plus ends. 
In contrast, in the chromosome-directed pathway, chromosomes induce 
MT assembly; but, the randomly oriented MTs require assistance of 
motor proteins to reorient properly so as to form the bipolar spindle.



$\bullet$ {\bf Chromosome condensation}



$\bullet$ {\bf Chromosome motility}

We now focus on chromosome motility, i.e., the mechanism of pulling the 
chromosomes towards the two opposite poles by the motors and microtubules 
which form the spindle. 

Several time-dependent quantities can be monitored during chromosome 
segregation to characterize the underlying dynamics. These include, 
for example, (a) separation between the spindle poles, (b) the distance 
between a kinetochore and the corresponding spindle pole, etc. 



\subsection{Chromosome segregation in prokaryotes: machines and mechanisms}

So far there are no convincing direct evidence for the existence of any 
mitotic spindle-like machinery in bacteria for post-replication 
segregation of chromosomes before cell division. However, there are 
more primitive motors which carry out chromosome segregation in 
bacteria. 



$\bullet$ {\bf FtsK: Chromosome segregation machine in E-coli} 



$\bullet$ {\bf SpoIIIE: Chromosome segregation machine in Bacillus subtilis} 

Normally {\it Bacillus subtilis}, a rod shaped bacterium, divides to 
two similar daughter cells. However, under some special circumstances, 
which leads to spore formation, a {\it Bacillus subtilis} divides 
asymmetrically into a small prespore and a larger mother cell. The 
translocation of the chromosome into the small prespore compartment 
is carried out by the motor protein SpoIIIE. Most of the fundamental 
questions on its operational mechanism are similar to those generic 
ones for helicases and translocases (including packaging motors for 
viral capsids). In particular, how does SpoIIIE, which anchors itself 
at the septum between the two compartments, translocate the DNA in 
the desired direction, namely, from the larger to the smaller 
compartment?
in the  



\subsection{Eukaryotic cell cytokinesis: machines and mechanisms}

The last step in cell division involves the physical separation of the 
contents (including the cytoplasm) of the parent cell into the two 
daughter cells. This process is called cytokinesis.

$\bullet$ {\bf Cytokinesis in animal cells} 

In dividing cells of animals and fungi, an actomyosin ring forms in the 
middle of the cell and its contraction generate the force required for 
cytokinesis. In other words, cytokinesis in animals and fungi are driven 
by a coordinated operation of the cell membrane and a cytoskeletal 
motor-filament system. How is the equatorial plane recognized by 
actin? Do the actin filaments nucleate in the equatorial plane of the 
cell itself or are the actin filaments transported there from elsewhere? 
How are the actin filaments and the myosin motors organized and how do 
they interact so as to generate the force responsible for furrow 
ingression? Do the actin filaments work like a tightening ``purse string'' 
or do the actin filaments work like a radially shrinking spokes of a 
bicycle wheel?



$\bullet$ {\bf Cytokinesis in plant cells} 

The mechanism of cytokinesis in plant cells is quite different from 
that in cells of animals and fungi.



\subsection{Prokaryotic cell cytokinesis: machines and mechanisms}

For bacterial cells, the mechanisms of locating the mid-cell and those 
of cytokinesis are now quite well understood.
An interesting finding of the experiments is that eukaryotic cells use 
microtubule-based machiery for chromosome movements and actin-based 
machinery for cytokinesis. In sharp contrast, prokaryotic cells do 
just the reverse.

\subsection{Division of mitochondria and chloroplats}

Mitochondria and plastids (including chlroplasts) have their own 
distinct genomes. The division of these organelles can also be 
broken up into the following majors events: (i) organellar 
chromosome separation, and (ii) organellokinesis, which would be 
the analogue of cytokinesis.


\subsection{Division of peroxisomes}


Just as mitochondria are the powerhouses of eukaryotic cells,
peroxisomes are often regarded as the garbage pail of the cell.
The division of peroxisomes can be divided roughly into three 
stages: (a) {\it elongation} of the peroxisome, (b) {\it constriction} 
of peroxisomal membrane, and (c) {\it fission} of the peroxisome.
In spite of fundamental differences in their structure and function, 
mitochondria and peroxisomes share quite a few componnts of the
machineries which drive their fission. For example, dynamin-like
proteins, which are involved in the fission of mitochondria, also
form the ring that is required for the fission of peroxisomes.

\subsection{Stem cell division}

Asymmetry in cell division have already provided some insights into 
the plausible mechanisms of asymmetric division of self-renewing 
stem cells in mammals. 


\subsection{Coordination between replication, recombination and segregation} 



\subsection{Role of motors in development}



\section{Miscellaneous natural and artificial molecular machines} 

%

\subsection{Prestin} 

Prestin is a transmembrane protein in the outer hair cells of the 
cochlea, an important constituent of the mammalian hearing organ. 
It is a unique machine in the sense that it is a {\it electro-mechanical 
transducer} that coverts electrical input directly into mechanical output.



\subsection{G-proteins} 



\subsection{Common structural features- AAA+ superfamily} 



\subsection{Common structural features- RecA-like domains} 



\section{Collective oscillations in active systems} 



\section{Molecular biomimetics- a bottom-up approach to nano-technology} 

Initially, technology was synonymous with macro-technology. The first
tools applied by primitive humans were, perhaps, wooden sticks and
stone blades. Later, as early civilizations started using levers,
pulleys and wheels for erecting enormous structures like pyramids.
Until nineteenth century, watch makers were, perhaps, the only people
working with small machines. Using magnifying glasses, they worked
with machines as small as $0.1 mm$. Micro-technology, dealing with
machines at the length scale of micrometers, was driven, in the
second half of the twentieth century, largely by the computer
miniaturization.

In 1959, Richard Feynman delivered a talk at a meeting of the American
Physical Society. In this talk, entitled ``{\it There's Plenty of Room
at the Bottom}'', Feynman drew attention of the scientific
community to the unlimited possibilities of manupilating and controlling
things on the scale of nano-meters. This famous talk is now accepted by
the majority of physicists as the defining moment of nano-technology.
In the same talk, in his characteristic style, Feynman noted that ''many
of the cells are very tiny, but they are very active, they manufacture
various substances, they walk around, they wiggle, and they do all kinds
of wonderful things- all on a very small scale''.

From the perspective of applied research, the natural molecular machines
opened up a new frontier of nano-technology. The miniaturization of
components for the fabrication of useful devices, which are essential
for modern technology, is currently being pursued by engineers
following mostly a top-down (from larger to smaller) approach. On the
other hand, an alternative approach, pursued mostly by chemists, is a
bottom-up (from smaller to larger) approach. We can benefit from Nature's 
billion year experience in nano-technolgy. We have given a long list of 
studies completed so far on  the architectural design of a nantural nanomachine, identification of its components and  monitoring the spatio-temporal coordination of these
components in the overall operation of the machine. The lessons learnt
from such investigations can then be utilized to design and synthesize
artificial nanomachines. In fact, the term
{\it biomimetics} has already become a popular buzzword; this field
deals with the design of artificial machines utilizing the principles
of natural bio-machines.  Even nano-robotics may no longer be a distant dream. 



\bigskip
\noindent {\bf ACKNOWLEDGMENTS}
\bigskip
A subset of these references served as the core content of a course 
on natural nanomachines which I have taught a few times at IIT Kanpur.  
It is my great pleasure to thank all my students whose stimulating and 
thought provoking questions have helped me in organizing a vast area 
of research topics in a sequence of coherent themes. I thank my 
research collaborators Aakash Basu, Meredith Betterton, Ashok Garai, 
Manoj Gopalakrishnan, Bindu Govindan, Philip Greulich, Katsuhiro 
Nishinari, Yasushi Okada, T. V. Ramakrishnan, Andreas Schsdschneider, 
Tripti Tripathi and Jian-Sheng Wang for enjoyable collaborations 
on molecular machines. I also thank Debasish Chaudhuri, Eric Galburt, 
Stephan Grill, Joe Howard, Frank J\"ulicher, Stefan Klumpp, Anatoly 
Kolomeisky, Benjamin Lindner, Reinhard Lipowsky, Roop Mallik, Gautam 
Menon, Alex Mogilner, Francois Nedelec, Sriram Ramaswamy, Krishanu Ray, 
Gunter Sch\"utz and Thomas Surrey for valuable discussions on molecular 
machines. 
My research on molecular machines have been supported by CSIR (India). 
I thank the Visitors Program of MPI-PKS Dresden (Germany), EMBL 
Heidelberg (Germany), Forschungszentrum J\"ulich (Germany), Theoretical 
Physics Department of University of Cologne (Germany), Physics department 
of NUS (Singapore), Physics department of IISc Bangalore (India), 
Department of Biological Sciences of TIFR Mumbai (India), National 
Center for Biological Sciences Bangalore (India) and Raman Research 
Institute Bangalore (India) for their generous hospitalities during 
the compilation of this resource letter over the last few years. 

\end{document}